\PassOptionsToPackage{dvipsnames,table}{xcolor}
\documentclass[authoryear,5pt]{elsarticle}

\usepackage{lineno,hyperref}

\usepackage{fullpage}

\usepackage{amsmath}
\usepackage{amsfonts}
\usepackage{caption}
\usepackage{subcaption}
\usepackage{soul}

\usepackage{natbib}
\usepackage{algorithm}
\usepackage{algpseudocode}%

\usepackage{booktabs}
\usepackage{epstopdf}

\usepackage{physics}
\usepackage[abs]{overpic}
\usepackage{tikz}
\usepackage{wasysym}
\usepackage{relsize}
\usepackage{pifont}
\usepackage{tabularx}
\usepackage{amssymb}
\usepackage{siunitx}
\usepackage{natbib}
\usepackage{amsmath}

\newcommand{\rout}[1]

\setcitestyle{authoryear,open={(},close={)}} 

\usepackage[dvipsnames]{xcolor}

\usepackage{graphicx}
\usepackage{float}
\usepackage[english]{babel}
\usepackage{hyperref}
\usepackage{amssymb}
\usepackage[normalem]{ulem}
\usepackage{amsmath}
\usepackage{amsthm}
\usepackage{amsfonts}
\usepackage{enumerate}
\usepackage{mathtools}

\usepackage{lineno}
\usepackage{bm}

\modulolinenumbers[5]

\journal{Elsevier}

\usepackage[dvipsnames,table]{xcolor} 

\graphicspath{{figure/}}

\usepackage{amssymb}

\errorcontextlines\maxdimen

\makeatletter
\newcommand*{\algrule}[1][\algorithmicindent]{\makebox[#1][l]{\hspace*{.5em}\vrule height .75\baselineskip depth .25\baselineskip}}%

\newcount\ALG@printindent@tempcnta
\def\ALG@printindent{%
    \ifnum \theALG@nested>0
        \ifx\ALG@text\ALG@x@notext
            \addvspace{-3pt}
        \else
            \unskip
            \ALG@printindent@tempcnta=1
            \loop
                \algrule[\csname ALG@ind@\the\ALG@printindent@tempcnta\endcsname]%
                \advance \ALG@printindent@tempcnta 1
            \ifnum \ALG@printindent@tempcnta<\numexpr\theALG@nested+1\relax
            \repeat
        \fi
    \fi
    }%
\usepackage{etoolbox}
\patchcmd{\ALG@doentity}{\noindent\hskip\ALG@tlm}{\ALG@printindent}{}{\errmessage{failed to patch}}
\makeatother
\begin{document}

\begin{frontmatter}

\title{Bifurcation reorganization governs snap-through pathways \\ in folded elastic ribbons}

\author{Weicheng Huang$^{1,\star}$, Qun Zhang$^{2}$, Bohan Zhang$^{3}$, Mingchao Liu$^{2,\star}$,}

\address{
$^{1}$School of Engineering, Newcastle University, Newcastle upon Tyne, NE1 7RU, UK\\
$^{2}$Department of Mechanical Engineering, University of Birmingham, Birmingham, B15 2TT, UK\\ 
$^{3}$Department of Engineering Mechanics, Northwestern Polytechnical University, Xi’an, 710072, China \\
$^{\star}$Corresponding authors: weicheng.huang@ncl.ac.uk (W.H.); m.liu.2@bham.ac.uk (M.L.).
}

\begin{abstract}
Snap-through in slender elastic structures is often viewed as a sudden transition between stable configurations, yet the underlying instability mechanisms and the direction of state switching can vary fundamentally.
{Different bifurcation pathways can lead to distinct transitions, while either of two competing metastable states may lose stability.
Here, we show that localized folds provide a geometric means of programming snap-through pathways and transition directions in elastic ribbons.
We study a ribbon with two folds placed symmetrically about its midpoint and demonstrate that varying the fold position reorganizes the bifurcation landscape.
Three distinct instability-mediated pathways are identified: a supercritical pitchfork followed by stable asymmetric deformation and subsequent saddle-node-induced snap-through, a subcritical pitchfork directly triggering asymmetric snap-through, and a saddle-node inducing symmetry-preserving snap-through.
Moreover, changing the fold position reverses which metastable branch loses stability, leading to both top-to-bottom and bottom-to-top transitions.
We show that these transitions are governed by three aspects of bifurcation reorganization: changes in pitchfork criticality, reversal of the ordering of pitchfork and saddle-node bifurcations, and switching of the destabilized metastable state.
Combining experiments, discrete differential geometry simulations, and numerical continuation, we identify six snap-through regimes arising from two transition directions and three instability-mediated pathways, and construct a global bifurcation map relating these regimes to fold position.
These results demonstrate that localized geometric features can reorganize the stability landscape to program not only how a structure snaps, but also which state snaps, providing a geometric strategy for controlling nonlinear transitions in multistable and morphing structures.}
\end{abstract}


\end{frontmatter}

\section{Introduction}

Snap-through instabilities in slender elastic structures provide a powerful mechanism for rapid and reversible shape transformation \citep{holmes2007snapping,yang2023morphing}.
{The mechanics of snap-through and buckling in curved beams and arches has a long history, dating back to classical studies of curved bars and shallow arches \citep{timoshenko1935buckling,timoshenko1961theory,schreyer1966buckling}, and has continued to receive attention in more recent theoretical and experimental investigations \citep{harvey2015coexisting}.}
By exploiting geometric nonlinearity and elastic energy storage, beams, ribbons, plates, and shells can transition abruptly between distant equilibrium configurations while remaining within the elastic regime \citep{stoll1994analysis,plaut2015snap,huang2024exploiting,huang2025tutorial,lu20262d}.
Such transitions are typically associated with nonlinear energy landscapes containing multiple stable and unstable equilibrium states, where changes in geometry, loading, or boundary conditions can induce large-amplitude configurational changes \citep{liu2021delayed,radisson2023elastic,yang2025snap}.
This behavior has been widely exploited in morphing structures \citep{liu2023snap,rahman2024shape}, fluidic devices \citep{gomez2017passive,jiao2020snap}, deployable systems \citep{lalisani2026snap}, mechanical metamaterials \citep{zhang2021architected,yue2026mechanical}, and soft robotics \citep{wang2023insect,park2025snap}, where fast actuation, multistable switching, and large deformation are desirable.
In these systems, snap-through is increasingly regarded not merely as a failure mode, but as a functional mechanism for programming mechanical responses \citep{yan2025snap,lu2026mechanical}.

For functional applications, however, determining the onset or threshold of snap-through alone is insufficient.
It is equally important to understand how the transition proceeds, since the deformation pathway influences the symmetry, dynamics, and resulting configuration of the snapped structure \citep{lu2023easy,de2024snap,feng2026twist}.
{The distinction between symmetric and asymmetric instability pathways has long been recognized in elastic stability theory.
For example, classical studies of compressed arches have established the distinction between symmetric and asymmetric snapping in shallow and deep arches, including the influence of geometric parameters on the relative stability of different deformation modes \citep{timoshenko1961theory,schreyer1966buckling}.
More generally, multiple equilibrium branches and instability modes may coexist in geometrically nonlinear structures, allowing transitions that either preserve the underlying symmetry or involve symmetry breaking \citep{radisson2023elastic,harvey2015coexisting}.
These distinct pathways can have important consequences for mechanical functionality.
Symmetry-preserving transitions may provide predictable morphing routes, whereas symmetry-breaking transitions can select among multiple asymmetric configurations and introduce sensitivity to imperfections or perturbations \citep{pontecorvo2013bistable,wang2024transient}.
Understanding and controlling the selection of snap-through pathways is therefore essential for programming reliable mechanical transitions.}

The selection of a snap-through pathway is closely related to the bifurcation structure of the underlying equilibrium landscape \citep{radisson2023elastic}.
{In geometrically symmetric systems, a pitchfork bifurcation provides a canonical mechanism for symmetry breaking, through which asymmetric equilibrium branches emerge from a symmetric branch \citep{gomez2017critical,yu2019bifurcations,huang2020shear}.
Importantly, the nature of this bifurcation determines how symmetry breaking develops.
A supercritical pitchfork can lead to continuous symmetry breaking through the emergence of stable asymmetric branches, which may subsequently lose stability and undergo snap-through \citep{liu2021delayed,radisson2023dynamic}.
In contrast, a subcritical pitchfork typically involves unstable asymmetric branches near the bifurcation point and can trigger an abrupt symmetry-breaking transition \citep{radisson2023dynamic,radisson2023elastic}.
A saddle-node bifurcation, on the other hand, corresponds to the disappearance of an equilibrium branch at a limit point and can directly trigger snap-through without preceding symmetry breaking \citep{yan2019static,simpkins2026snap,bhattacharyya2026phenomenology}.
Consequently, both the criticality of a pitchfork bifurcation and the relative ordering of competing pitchfork and saddle-node bifurcations can influence the resulting transition pathway.
The competition between symmetric and asymmetric instabilities, and the dependence of the governing instability mode on structural geometry, have long been studied in classical arch mechanics \citep{timoshenko1961theory,schreyer1966buckling}, while geometric control of pitchfork--saddle-node ordering has also been demonstrated in more recent programmable multistable systems \citep{zhang2025achieving}.
These studies establish the importance of bifurcation structure in determining snap-through behavior and motivate further investigation into how different instability mechanisms can be systematically organized through structural design.

Localized folds and creases provide an appealing geometric route for controlling the stability of slender structures.}
In origami-inspired systems, creases are commonly employed to prescribe folding kinematics, introduce localized compliance, and enable programmable shape transformations \citep{dias2014non,misseroni2024origami}.
Beyond classical origami settings, folded and creased ribbons offer a rich mechanical platform in which localized geometric features interact with global bending, twisting, and buckling modes \citep{jules2019local,marzin2025augmented}.
For example, creased annuli exhibit tunable bistability and looping behavior \citep{yu2023continuous}, while kinks and creases in meta-ribbons enable programmable folding sequences and reconfigurable morphing architectures \citep{huang2024integration}.
{More recently, localized folds have been shown to reshape the stability landscape of elastic ribbons by shifting, biasing, or suppressing pitchfork bifurcations, thereby controlling symmetry-breaking deformation under transverse displacement \citep{huang2026programmable}.
Unlike approaches based on distributed actuation or changes in global structural geometry, localized fold placement provides a passive geometric means of modifying the coupled bending and twisting response of a three-dimensional ribbon.
This raises a broader question: can the placement of localized folds systematically reorganize not only individual instability thresholds, but also the criticality and ordering of competing bifurcations, thereby controlling both the pathway and direction of snap-through between metastable states?}

Here, we address this question by investigating pre-compressed elastic ribbons containing a pair of symmetrically positioned localized folds, as illustrated in Fig.~\ref{fig:overview}.
{We identify three distinct instability-mediated pathways for top-to-bottom snap-through: a supercritical pitchfork followed by stable asymmetric deformation and subsequent saddle-node-induced snapping, a subcritical pitchfork directly triggering asymmetric snapping, and a saddle-node triggering symmetric snapping.
By varying the fold position, we demonstrate how these pathways are connected through changes in pitchfork criticality and the relative ordering of pitchfork and saddle-node bifurcations.
Further changes in fold position switch the destabilized metastable state, reversing the snap-through direction from top-to-bottom to bottom-to-top.
The three instability-mediated pathways also emerge in the reverse direction, giving rise to six distinct snap-through regimes.
Combining experiments, discrete modeling, dynamic simulations, and numerical continuation, we establish a global bifurcation map that relates the critical configurations and loading thresholds to fold position.
Together, these results demonstrate how localized geometric folds can reorganize the bifurcation landscape of multistable elastic structures to program the symmetry, pathway, and direction of snap-through.}

The remainder of this article is organized as follows.
Section~\ref{sec:Experiment} introduces the physical problem and experimental platform.
Section~\ref{sec:simulation} presents the discrete model and stability analysis.
{Section~\ref{sec:results} first examines three top-to-bottom snap-through pathways, then investigates their transitions with fold position and the emergence of corresponding bottom-to-top pathways, before establishing the global organization of the six regimes.}
Finally, conclusions are presented in Section~\ref{sec:discussion}.

\section{Problem setup and experimental platform}
\label{sec:Experiment}

\begin{figure}[!h]
    \centering
    \includegraphics[width=\textwidth]{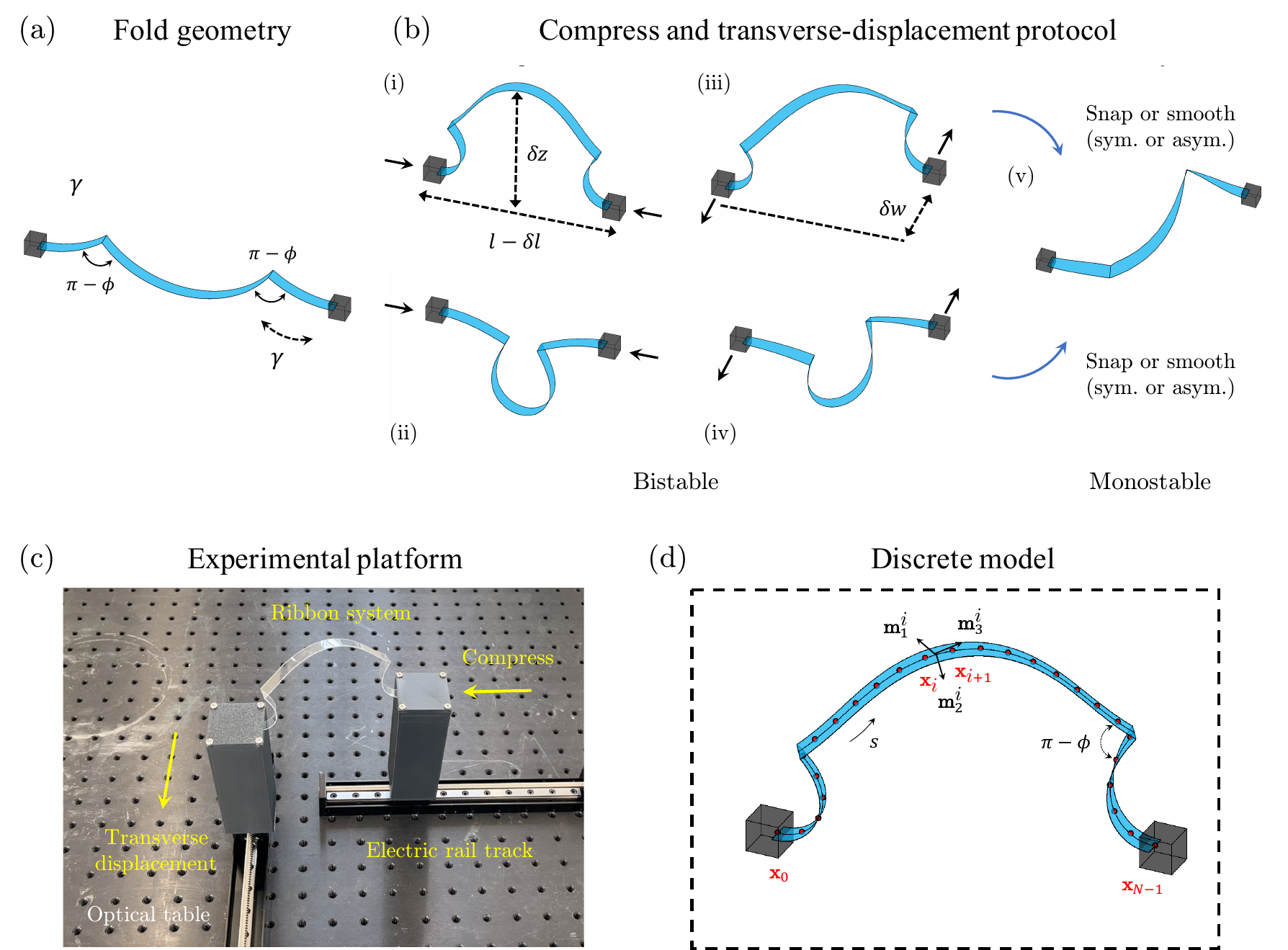}
\caption{
    Snap-through of folded elastic ribbons: physical problem, experimental setup, and discrete model.
    (a) Initial uncompressed ribbon containing two symmetrically positioned localized folds, characterized by the fold angle $\phi$ and fold position $\gamma$.
    (b) Loading protocol. Longitudinal compression with end-shortening $\delta l$ produces two stable pre-compressed configurations: (i) an upper state and (ii) a lower state, with opposite signs of the midpoint displacement. A subsequent transverse displacement $\delta w$ is applied while maintaining the end-shortening, driving the system from a bistable to a monostable state. Depending on the fold position, either pre-compressed configuration may lose stability, leading to top-to-bottom or bottom-to-top snap-through through a symmetric or asymmetric pathway.
    {(c) Experimental platform, consisting of a folded ribbon mounted between two actuators. The right actuator controls longitudinal compression, while the left actuator imposes transverse displacement.}
    (d) Discrete differential geometry (DDG) representation of the ribbon, with its centerline discretized into vertices $\mathbf{x}_i$ connected by elastic segments and equipped with discrete material frames $\{\mathbf{m}_1^i,\mathbf{m}_2^i,\mathbf{m}_3^i\}$ to describe bending and twisting deformation. Localized folds are represented through prescribed intrinsic curvature.
    }
    \label{fig:overview}
\end{figure}

We consider a thin, foldable elastic ribbon subjected to prescribed boundary displacements, as illustrated in Fig.~\ref{fig:overview}.
The ribbon has initial length $l$, width $w$, and thickness $b$, satisfying the slenderness condition $l \gg w \gg b$.
A pair of localized folds is introduced symmetrically about the ribbon midpoint, with the folds located at $s=\gamma$ and $s=l-\gamma$ and characterized by a prescribed fold angle $\phi$.
The normalized fold position is defined as $\Gamma=\gamma/l$.
The fold position and fold angle provide geometric design parameters for programming the stability landscape of the ribbon.

The mechanical response is studied through a three-stage loading protocol.
Initially, the ribbon is in an uncompressed folded configuration containing the prescribed localized folds, as shown in Fig.~\ref{fig:overview}(a).
The ribbon is then subjected to longitudinal compression, reducing the distance between the two clamped ends from $l$ to $l-\delta l$.
This compression deforms the folded ribbon into pre-compressed arched configurations with characteristic midpoint height $\delta z$.
At this stage, two stable equilibrium configurations coexist: an upper pre-compressed state, shown in Fig.~\ref{fig:overview}(b)(i), and a lower pre-compressed state, shown in Fig.~\ref{fig:overview}(b)(ii), distinguished by the sign of the midpoint height.
Subsequently, a transverse displacement $\delta w$ is applied at one end while the end-to-end distance $l-\delta l$ is maintained.
The transverse displacement serves as the primary control parameter in the stability analysis.
{As the transverse displacement increases, the stability of the two pre-compressed configurations evolves, and the system may undergo symmetry breaking and snap-through through different instability pathways.
Depending on the fold position, either the upper or lower metastable state can lose stability, leading to top-to-bottom or bottom-to-top snap-through, respectively.
Moreover, the transition may remain symmetric or involve an asymmetric deformation pathway, depending on the sequence and nature of the underlying bifurcations.
The central question of this work is how localized fold placement controls the bifurcation structure and thereby determines the pathway, symmetry, and direction of snap-through.}

The experiments are conducted on a custom-designed platform for imposing prescribed boundary displacements, as illustrated in Fig.~\ref{fig:overview}(c).
The system consists of two programmable actuators mounted on an electric linear rail, allowing accurate control of the longitudinal shortening $\delta l$ and transverse displacement $\delta w$.
The setup is placed on an optical table to ensure mechanical stability, and a camera is used to record the evolution of the ribbon configurations during deformation.
{The right actuator applies longitudinal compression, while the left actuator imposes transverse displacement.}
The ribbon specimens are fabricated from elastic polyethylene terephthalate (PET) sheets with length $l=200$~mm, width $w=5.0$~mm, and thickness $b=0.25$~mm.
Localized folds are introduced at prescribed symmetric positions, $s=\gamma$ and $s=l-\gamma$, {following the fabrication strategy used for elastic creased structures in \citep{yu2023continuous}.
Specifically, the ribbon is heat-set against a prescribed geometric template to introduce the target fold geometry into its stress-free configuration.
The fold position and fold angle are determined by the geometry of the template.
The fold angle $\pi-\phi$ is measured before and after the mechanical tests, with no noticeable change observed, indicating that the fold geometry remains essentially unchanged throughout the experiments.
This is consistent with the relatively high stiffness of the folded region compared with the surrounding flexible portions of the ribbon.
The nominal fold angle is $\phi=80^{\circ}$, with a measured variation of $\pm 2^{\circ}$, and three specimens are tested for each configuration, with the snap-through pathway being reproducible across all trials.
The uncertainty in the fold position is within $\pm 1$~mm for a ribbon of total length $l=200$~mm, corresponding to a normalized fold-position uncertainty of $\pm 0.005$.

The deformation process was recorded using an iPhone camera.
Although the loading was quasi-static, the snap-through transition is transient, and the camera frame rate was sufficient to capture the overall snap-through pathway and the associated sequence of configurational changes.}

\section{Discrete model and stability analysis}
\label{sec:simulation}

We build on an established discrete differential geometry (DDG) framework to model the nonlinear deformation and stability of slender folded ribbons with localized out-of-plane creases \citep{huang2025tutorial}. The ribbon is represented as an anisotropic slender body capable of undergoing large rotations and geometrically nonlinear deformation, while the prescribed folds are incorporated through localized curvature discontinuities along the centerline. Building on previous implementations for folded ribbons \citep{huang2024integration,huang2026programmable}, the present framework combines numerical continuation, eigenvalue-based stability analysis and dynamic simulation to track the competing pitchfork and saddle-node bifurcations in a two-fold geometry and resolve the associated post-instability response.

The ribbon is modeled as a slender body of length $l$, width $w$, and thickness $b$, satisfying $l \gg w \gg b$. Under this geometric separation of scales, an anisotropic rod idealization provides an efficient description of the nonlinear mechanics of narrow strips and folded ribbons \citep{yu2019bifurcations,huang2020shear,huang2024exploiting,huang2025tutorial}. The cross-sectional area is $A=wb$, the principal second moments of area are $I_{1}=wb^{3}/12$ and $I_{2}=bw^{3}/12$, and the torsional constant is approximated as $J=wb^{3}/3$. The ribbon is made of an elastic material with Young's modulus $E$, shear modulus $G$, and density $\rho$.

{We note that the term ribbon in this work refers to the physical geometry of the slender strip, whereas the numerical formulation is based on an anisotropic elastic rod model rather than a classical Sadowsky- or Wunderlich-type ribbon theory.
The anisotropic rod approximation is appropriate for sufficiently slender strips satisfying $l\gg w\gg b$ \citep{audoly2021one}. This condition is well satisfied by the present specimens, with $l=200$ mm, $w=5$ mm, and $b=0.25$ mm, corresponding to $l/w=40$ and $w/b=20$.
Unlike classical developable-ribbon theories, which explicitly account for the geometric constraint of developability, the anisotropic rod formulation describes the ribbon through its centerline and cross-sectional orientation, with anisotropic bending and twisting stiffnesses.
Although this approximation does not explicitly enforce developability, it provides a computationally efficient framework for describing the three-dimensional bending, twisting, and bifurcation behavior considered here \citep{huang2020shear}.}

\paragraph{Kinematics} Referring to Fig.~\ref{fig:overview}(d), the ribbon centerline is discretized into $N$ nodes,
\begin{equation}
\left[ \mathbf{x}_{0}, \mathbf{x}_{1}, \ldots, \mathbf{x}_{N-1} \right],
\end{equation}
which define $N-1$ edge vectors
\begin{equation}
\left[ \mathbf{e}^{0}, \mathbf{e}^{1}, \ldots, \mathbf{e}^{N-2} \right],
\end{equation}
with
\begin{equation}
\mathbf{e}^{i} = \mathbf{x}_{i+1} - \mathbf{x}_{i}.
\end{equation}
To describe the orientation of each edge, we associate with edge $i$ an orthonormal reference frame
\begin{equation}
\left\{\mathbf{d}^{i}_{1}, \mathbf{d}^{i}_{2}, \mathbf{d}^{i}_{3}\right\},
\end{equation}
and a material frame
\begin{equation}
\left\{\mathbf{m}^{i}_{1}, \mathbf{m}^{i}_{2}, \mathbf{m}^{i}_{3}\right\}.
\end{equation}
Both frames are adapted to the centerline, so that their third director coincides with the unit tangent,
\begin{equation}
\mathbf{d}^{i}_{3} \equiv \mathbf{m}^{i}_{3}
=
\frac{\mathbf{e}^{i}}{\|\mathbf{e}^{i}\|}.
\end{equation}
For a ribbon, the first material director $\mathbf{m}^{i}_{1}$ is chosen as the local surface normal, and the second director $\mathbf{m}^{i}_{2}$ spans the width direction,
\begin{equation}
\mathbf{m}^{i}_{2} = \mathbf{m}^{i}_{3} \times \mathbf{m}^{i}_{1}.
\end{equation}
The reference frame is updated by parallel transport along the centerline \citep{bergou2008discrete,bergou2010discrete}, while the material frame is obtained from the reference frame through a scalar twist angle $\theta^{i}$ on each edge. The discrete degrees of freedom are therefore collected into
\begin{equation}
\mathbf{q}
=
\left[
\mathbf{x}_{0},\theta^{0},
\mathbf{x}_{1},\theta^{1},
\ldots,
\mathbf{x}_{N-2},\theta^{N-2},
\mathbf{x}_{N-1}
\right]^{\mathrm T},
\end{equation}
which contains $3N+(N-1)$ unknowns.

\paragraph{Elastic energy} The elastic deformation of the ribbon consists of stretching, bending, and twisting \citep{huang2025tutorial}. The axial strain associated with edge $i$ is defined by
\begin{equation}
\varepsilon^{i}
=
\frac{\|\mathbf{e}^{i}\|}{\|\hat{\mathbf{e}}^{i}\|}-1,
\label{eq:stretching_strain}
\end{equation}
where a hat denotes a quantity evaluated in the stress-free reference configuration, and $\|\cdot\|$ is the Euclidean norm.
Bending is measured at the internal node $\mathbf{x}_{i}$, $i=1,\ldots,N-2$, through the curvature binormal
\begin{equation}
(\boldsymbol{\kappa}\mathbf{b})_{i}
=
\frac{
2\,\mathbf{e}^{i-1}\times\mathbf{e}^{i}
}{
\|\mathbf{e}^{i-1}\|\,\|\mathbf{e}^{i}\|
+
\mathbf{e}^{i-1}\cdot\mathbf{e}^{i}
}.
\end{equation}
The two material curvatures are then obtained by projecting the curvature binormal onto the averaged material frame,
\begin{equation}
\kappa_{1,i}
=
\frac{1}{2}
\frac{
\left(\mathbf{m}_{2}^{i-1}+\mathbf{m}_{2}^{i}\right)\cdot(\boldsymbol{\kappa}\mathbf{b})_{i}
}{
\Delta l_{i}
},
\end{equation}
\begin{equation}
\kappa_{2,i}
=
-\frac{1}{2}
\frac{
\left(\mathbf{m}_{1}^{i-1}+\mathbf{m}_{1}^{i}\right)\cdot(\boldsymbol{\kappa}\mathbf{b})_{i}
}{
\Delta l_{i}
},
\label{eq:bending_strain}
\end{equation}
where
\begin{equation}
\Delta l_{i}
=
\frac{\|\mathbf{e}^{i-1}\|+\|\mathbf{e}^{i}\|}{2}
\end{equation}
is the Voronoi length associated with node $i$. Here, $\kappa_{1,i}$ and $\kappa_{2,i}$ denote the curvatures about the $\mathbf{m}_{1}$- and $\mathbf{m}_{2}$-axes, respectively.
The discrete twist strain at node $i$ is defined as \citep{bergou2008discrete,bergou2010discrete,huang2025tutorial},
\begin{equation}
\kappa_{3,i}
=
\frac{
\theta^{i}-\theta^{i-1}+m^{\mathrm{ref}}_{i}
}{
\Delta l_{i}
},
\qquad i=1,\ldots,N-2,
\label{eq:twisting_curvature}
\end{equation}
where $m^{\mathrm{ref}}_{i}$ is the reference twist induced by parallel transport of the reference frame.
The total elastic energy is written as
\begin{equation}
E^{\mathrm{ela}}
=
E^{\mathrm{s}}
+
E^{\mathrm{b}}_{1}
+
E^{\mathrm{b}}_{2}
+
E^{\mathrm{t}},
\end{equation}
with
\begin{equation}
\begin{aligned}
E^{\mathrm{s}}
&=
\sum_{i=0}^{N-2}
\frac{1}{2}\,EA_{i}\,(\varepsilon^{i})^{2}\,\|\hat{\mathbf{e}}^{i}\|, \\
E^{\mathrm{b}}_{1}
&=
\sum_{i=1}^{N-2}
\frac{1}{2}\,EI_{1,i}\,(\kappa_{1,i}-\hat{\kappa}_{1,i})^{2}\,\Delta \hat{l}_{i}, \\
E^{\mathrm{b}}_{2}
&=
\sum_{i=1}^{N-2}
\frac{1}{2}\,EI_{2,i}\,(\kappa_{2,i}-\hat{\kappa}_{2,i})^{2}\,\Delta \hat{l}_{i}, \\
E^{\mathrm{t}}
&=
\sum_{i=1}^{N-2}
\frac{1}{2}\,GJ_{i}\,(\kappa_{3,i}-\hat{\kappa}_{3,i})^{2}\,\Delta \hat{l}_{i}.
\end{aligned}
\label{eq:totalEnergy}
\end{equation}
Here, $EA_{i}$ is the axial stiffness, $EI_{1,i}$ and $EI_{2,i}$ are the two bending stiffnesses, and $GJ_{i}$ is the twisting stiffness. For slender ribbons, the axial stiffness is typically much larger than the bending and twisting stiffnesses, so the stretching energy effectively acts as a penalty term that enforces near-inextensibility of the centerline \citep{huang2025tutorial}.

\paragraph{Abrupt folds}
A key advantage of the present discrete formulation is that geometric discontinuities can be introduced naturally at the nodal level. If a crease with prescribed turning angle $\phi_{i}$ is placed at node $\mathbf{x}_{i}$, its effect is represented through an imposed intrinsic curvature \citep{huang2024integration}
\begin{equation}
\hat{\kappa}_{2,i}
=
2\,\frac{\tan(\phi_{i}/2)}{\Delta l_{i}}.
\end{equation}
To model a sharp fold, the corresponding bending stiffness $EI_{2,i}$ is chosen sufficiently large, so that the constraint $\kappa_{2,i}  =  \hat{\kappa}_{2,i}$ is strongly enforced \citep{huang2026programmable}. In this way, the discrete model can accommodate both smooth elastic deformation and localized fold kinematics within a unified energy-based framework.
{The fold is therefore represented by a single node with a finite prescribed turning angle $\phi_i$, which remains localized for any $N$. When $N$ is too small, however, every node becomes a coarse kink and the polyline can no longer approximate a smooth continuous curve, so a sufficiently fine mesh is required.}

\paragraph{Boundary condition} To impose the clamped--clamped boundary condition, the first two nodes \( \{ \mathbf{x}_{0}, \mathbf{x}_{1} \} \), the last two nodes \( \{ \mathbf{x}_{N-2}, \mathbf{x}_{N-1} \} \), and the associated rotational degrees of freedom \( \{ \theta^{0}, \theta^{N-2} \} \) are prescribed and updated consistently with the boundary constraints.

\paragraph{Static arc-length continuation}
The equilibrium path of folded ribbons generally contains limit points associated with snap-through and snap-back instabilities. In such cases, conventional load-controlled Newton iterations fail once the tangent stiffness matrix becomes singular. To trace the complete equilibrium branch, including unstable segments, we employ the modified generalized displacement control method (MGDCM), which is a robust arc-length-type continuation scheme \citep{leon2014effect,liu2017Nonlinear}.
The nonlinear equilibrium equation is
\begin{equation}
\mathbf{R}(\mathbf{q},\lambda)
=
\lambda \mathbf{F}^{\mathrm{ext}}
+
\mathbf{F}^{\mathrm{ela}}(\mathbf{q})
=
\mathbf{0},
\end{equation}
where $\mathbf{q}$ is the generalized coordinate vector, $\mathbf{F}^{\mathrm{ela}}$ is the internal elastic force vector, $\mathbf{F}^{\mathrm{ext}}$ is the reference external load vector, and $\lambda$ is the load factor.
Note that the internal elastic force vector is obtained from the variation of the total elastic energy,
\begin{equation}
\mathbf{F}^{\mathrm{ela}}
=
-
\frac{\partial E^{\mathrm{ela}}}{\partial \mathbf{q}}.
\end{equation}
At iteration $j$ of load step $k$, the linearized equilibrium equation is
\begin{equation}
\mathbb{K}^{k}_{j-1}\,\Delta \mathbf{q}^{k}_{j}
=
\Delta\lambda^{k}_{j}\,\mathbf{F}^{\mathrm{ext}}
+
\mathbf{R}^{k}_{j-1},
\label{eq:linearized_equilibrium}
\end{equation}
where the tangent stiffness matrix is defined by
\begin{equation}
\mathbb{K}
=
\frac{\partial^2 E^{\mathrm{ela}}}{\partial \mathbf{q}^2},
\end{equation}
and the residual vector is
\begin{equation}
\mathbf{R}^{k}_{j-1}
=
(\lambda^{k}_{j-1}) \mathbf{F}^{\mathrm{ext}}
+
\mathbf{F}^{\mathrm{ela}}(\mathbf{q}^{k}_{j-1}).
\end{equation}
The solution is updated according to
\begin{equation}
\mathbf{q}^{k}_{j}
=
\mathbf{q}^{k}_{j-1}
+
\Delta \mathbf{q}^{k}_{j},
\qquad
\lambda^{k}_{j}
=
\lambda^{k}_{j-1}
+
\Delta\lambda^{k}_{j}.
\end{equation}
Since both $\Delta \mathbf{q}^{k}_{j}$ and $\Delta\lambda^{k}_{j}$ are unknown, Eq.~\eqref{eq:linearized_equilibrium} is supplemented by a scalar constraint equation,
\begin{equation}
\mathbf{a}^{k}_{j}\cdot \Delta \mathbf{q}^{k}_{j}
+
b^{k}_{j}\,\Delta\lambda^{k}_{j}
=
c^{k}_{j}.
\end{equation}
The displacement increment is decomposed as
\begin{equation}
\Delta \mathbf{q}^{k}_{j}
=
(\Delta \mathbf{q}^{\mathrm{r}})^{k}_{j}
+
\Delta\lambda^{k}_{j}\,
(\Delta \mathbf{q}^{\mathrm{ext}})^{k}_{j},
\end{equation}
where $(\Delta \mathbf{q}^{\mathrm{r}})^{k}_{j}$ and $(\Delta \mathbf{q}^{\mathrm{ext}})^{k}_{j}$ satisfy
\begin{equation}
\mathbb{K}^{k}_{j-1}\,
(\Delta \mathbf{q}^{\mathrm{r}})^{k}_{j}
=
\mathbf{R}^{k}_{j-1},
\end{equation}
\begin{equation}
\mathbb{K}^{k}_{j-1}\,
(\Delta \mathbf{q}^{\mathrm{ext}})^{k}_{j}
=
\mathbf{F}^{\mathrm{ext}}.
\end{equation}
Substitution into the constraint equation gives
\begin{equation}
\Delta\lambda^{k}_{j}
=
\frac{
c^{k}_{j}
-
\mathbf{a}^{k}_{j}\cdot(\Delta \mathbf{q}^{\mathrm{r}})^{k}_{j}
}{
\mathbf{a}^{k}_{j}\cdot(\Delta \mathbf{q}^{\mathrm{ext}})^{k}_{j}
+
b^{k}_{j}
}.
\end{equation}
Based on the above formulation, the load factor increment is updated as \citep{leon2014effect,liu2017Nonlinear},
\begin{equation}
\Delta \lambda^{k}_{j}=
\begin{cases}
\Delta \bar{\lambda}, & k=1,\; j=1, \\[6pt]
-\dfrac{
(\Delta \mathbf{q}^{\mathrm{ext}}) ^{1}_{1}  
\cdot
(\Delta \mathbf{q}^{\mathrm{r}}) ^{1}_{j}
}
{
(\Delta \mathbf{q}^{\mathrm{ext}}) ^{1}_{1} 
\cdot
(\Delta \mathbf{q}^{\mathrm{ext}}) ^{1}_{j} 
},
& k=1,\; j>1, \\[10pt]
\pm \Delta \bar{\lambda} 
\sqrt{
\dfrac{
(\Delta \mathbf{q}^{\mathrm{ext}}) ^{1}_{1}  
\cdot
(\Delta \mathbf{q}^{\mathrm{ext}}) ^{1}_{1}
}
{
(\Delta \mathbf{q}^{\mathrm{ext}}) ^{k}_{1} 
\cdot
(\Delta \mathbf{q}^{\mathrm{ext}}) ^{k}_{1} 
}
}
, & k>1,\; j=1 ,\\[10pt]
-\dfrac{
(\Delta \mathbf{q}^{\mathrm{ext}}) ^{k}_{1}  
\cdot
(\Delta \mathbf{q}^{\mathrm{r}}) ^{k}_{j}
}
{
(\Delta \mathbf{q}^{\mathrm{ext}}) ^{k}_{1} 
\cdot
(\Delta \mathbf{q}^{\mathrm{ext}}) ^{k}_{j} 
},
& k>1,\; j>1. \\[10pt]
\end{cases}
\end{equation}
Here, the sign of the load factor increment is determined according to the direction of the displacement increment between two consecutive load steps, i.e.,
\begin{equation}
\mathrm{sgn}\!\left(
(\Delta \mathbf{q}^{\mathrm{ext}})^{k-1}_{1}
\cdot
(\Delta \mathbf{q}^{\mathrm{ext}})^{k}_{1}
\right).
\end{equation}
The parameter $\Delta \bar{\lambda}$ represents the prescribed initial load increment.
Local stability is determined from the tangent stiffness matrix. An equilibrium configuration is stable when all eigenvalues remain positive, while a bifurcation point is signaled by the vanishing of one eigenvalue.
The corresponding eigenvector supplies a natural perturbation direction for branch switching.
This procedure allows the solver to systematically trace connected equilibrium branches and their bifurcation structure.

\paragraph{Dynamic simulation}
In addition to static continuation, we employ dynamic relaxation to compute stable post-buckled configurations and to capture rapid transitions between equilibria \citep{huang2025tutorial}. The governing equation is written as
\begin{equation}
\mathbb{M}\,\ddot{\mathbf{q}}
+
\mathbb{C}\,\dot{\mathbf{q}}
+
\mathbb{K}\, {\mathbf{q}}
=
\mathbf{F}^{\mathrm{ext}}
\label{eq:dynamic_equilibrium}
\end{equation}
where $\mathbb{M}$ is the diagonal mass matrix  and $\mathbb{C} = \mu \mathbb{M}$ is a viscous damping matrix.
Equation~\eqref{eq:dynamic_equilibrium} is integrated in time using the implicit Euler scheme. Advancing from $t_{k}$ to $t_{k+1}=t_{k}+\delta t$, we write
\begin{equation}
\mathbb{M}\,\ddot{\mathbf{q}} (t_{k+1})
+
\mathbb{C}\,\dot{\mathbf{q}}(t_{k+1})
+
\mathbb{K}\,{\mathbf{q}} (t_{k+1})
=
 \mathbf{F}^{\mathrm{ext}}  (t_{k+1}),
\end{equation}
\begin{equation}
\mathbf{q}  (t_{k+1})
=
\mathbf{q} (t_{k})
+
\delta t\,\dot{\mathbf{q}} (t_{k+1}),
\end{equation}
\begin{equation}
\dot{\mathbf{q}} (t_{k+1})
=
\dot{\mathbf{q}} (t_{k})
+
\delta t\,\ddot{\mathbf{q}} (t_{k+1}).
\end{equation}
The inertia and damping can capture the transient response and drive the system toward a stable equilibrium state.
This strategy is particularly effective for transient problems involving severe nonlinearity, snap events, and abrupt changes in configuration.

\section{Geometric reorganization of bifurcations and snap-through pathways} 
\label{sec:results}

In this section, we present combined numerical and experimental results to characterize the configurational evolution and snap-through pathways of folded elastic ribbons under transverse displacement.
We consider a ribbon with two symmetrically distributed folds located at $\{\Gamma,1-\Gamma\}=\{\gamma/l,1-\gamma/l\}$.
Our main objective is to understand how fold position controls the nature and sequence of competing bifurcations and thereby determines the symmetry, pathway, and direction of snap-through.
Throughout this section, the axial compression between the two clamped ends is normalized as $\Delta L=\delta l/l$, and the transverse displacement is defined as $\Delta W=\delta w/l$.
Unless otherwise specified, the compression is fixed at $\Delta L=0.5$, and $\Delta W$ is taken as the primary bifurcation parameter.
The ribbon deformation is characterized by the normalized midpoint height $\Delta Z=\delta z/l$.
We examine the evolution of both the upper and lower pre-compressed equilibrium states, denoted by $\mathcal{C}^{P}$ and $\mathcal{C}^{M}$, respectively, and investigate how their loss of stability gives rise to top-to-bottom or bottom-to-top snap-through.
To quantify the degree of symmetry during deformation, we define the asymmetry measure $\mathcal{A}$ as
{
\begin{equation}
\mathcal{A}
=
\int_{0}^{{l}/{2}} || \kappa_{2}(s/l) - \kappa_{2}(1-s/l) || ds.
\end{equation}
A vanishing value of $\mathcal{A}$ indicates reflection symmetry of the curvature field, whereas a nonzero value quantifies its asymmetry. In the present deformation pathways, this measure is used together with the ribbon configurations to identify symmetry breaking.
}

In the numerical simulations, the ribbon is taken to have length $l=1000$ mm, width $w=20$ mm, and thickness $b=1$ mm, corresponding to $l/w=50$ and $w/b=20$.
{These geometric ratios are close to those used in the experiments, where $l/w=40$ and $w/b=20$.}
{The material parameters used in the numerical simulations are Young's modulus $E=1$~GPa, Poisson's ratio $\nu=0.5$, and density $\rho=1000$~kg/m$^3$}.
For the static continuation, an adaptive step size is used.
In the dynamic simulations, the transverse displacement is applied at a constant rate $\delta \dot{w}=0.01~\mathrm{m/s}$, which is sufficiently small to ensure a quasi-static loading process.
Unless otherwise specified, the ribbon is discretized with $N=201$ nodes, with a time step of $\delta t=1~\mathrm{ms}$ and a mass-proportional damping coefficient of $\mu=0.1$.
{At each fold node, the bending stiffness is increased to $K_f = 200\,EI_2$ to enforce the prescribed intrinsic curvature $\hat{\kappa}_{2,i}$.
To trigger branch selection when symmetry breaking occurs, a small imperfection is introduced as a slight rotation of the left clamped edge, with amplitude $\delta\alpha = 0.001$.
The mesh-convergence study is presented in~\ref{sec:AppendixConvergent}.}

\subsection{Three distinct top-to-bottom snap-through pathways}

{We first investigate three representative fold positions, $\Gamma=0.08$, $0.14$, and $0.21$, for which the ribbon undergoes top-to-bottom snap-through from the plus state $\mathcal{C}^{P}$ to the minus state $\mathcal{C}^{M}$.
Although all three cases involve the same transition direction, they exhibit distinct instability-mediated pathways governed by the nature and relative ordering of pitchfork and saddle-node bifurcations.
Specifically, the transition may occur through (i) a supercritical pitchfork followed by stable asymmetric deformation and subsequent saddle-node-induced snapping, (ii) a subcritical pitchfork directly triggering asymmetric snapping, or (iii) a saddle-node triggering symmetric snapping before symmetry breaking occurs.
These three cases demonstrate how different bifurcation sequences can lead to distinct snap-through pathways even when the initial and final metastable states remain the same.}


\begin{figure}[!b]
    \centering
    \includegraphics[width=\textwidth]{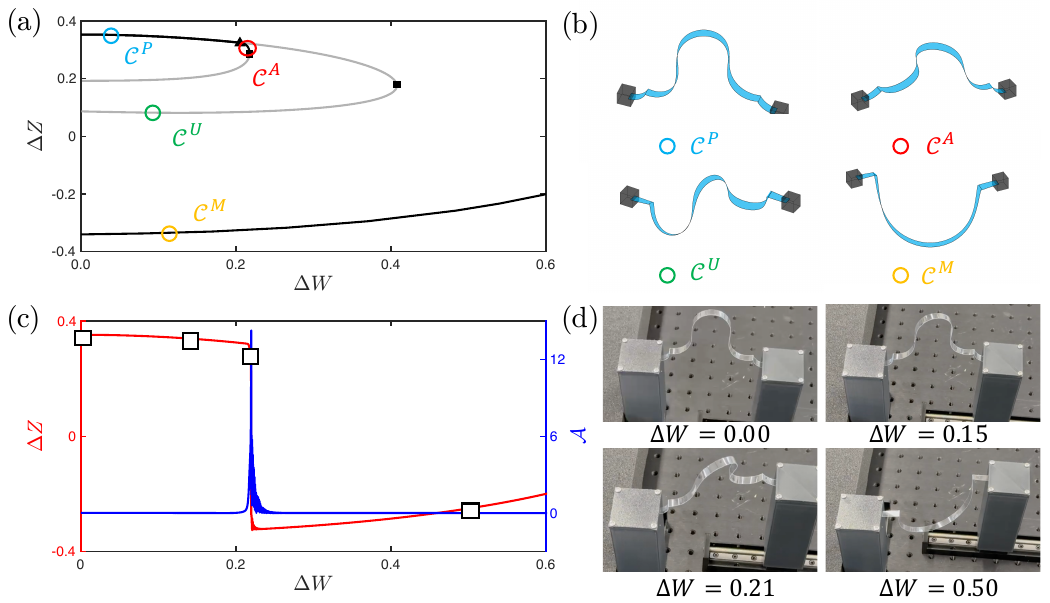}
    \caption{
    {Supercritical pitchfork bifurcation followed by evolution along a stable asymmetric branch and subsequent saddle-node-induced asymmetric top-to-bottom snap-through for fold position $\Gamma=16/200=0.08$.
    (a) Static bifurcation diagram showing the normalized midpoint height $\Delta Z$ as a function of transverse displacement $\Delta W$.
    The plus state $\mathcal{C}^{P}$ first undergoes a supercritical pitchfork bifurcation, giving rise to a stable asymmetric branch along which the ribbon evolves before reaching a saddle-node bifurcation that triggers snap-through.
    (b) Representative numerical configurations associated with the plus state $\mathcal{C}^{P}$, asymmetric state $\mathcal{C}^{A}$, unstable state $\mathcal{C}^{U}$, and minus state $\mathcal{C}^{M}$.
    The colored circles in (a) mark the corresponding equilibrium configurations shown in (b).
    (c) Dynamic evolution of $\Delta Z$ and the asymmetry measure $\mathcal{A}$.
    The growth of $\mathcal{A}$ demonstrates the asymmetric character of the transition.
    The open squares mark the selected stages corresponding to the experimental snapshots in (d).
    (d) Representative experimental snapshots during the transition.}
    }
    \label{fig:snap_16}
\end{figure}

\paragraph{Supercritical pitchfork followed by asymmetric snap-through} 
{We first consider the case with fold position $\Gamma=16/200=0.08$, as shown in Fig.~\ref{fig:snap_16}.
The static bifurcation diagram in Fig.~\ref{fig:snap_16}(a) shows that, with increasing transverse displacement $\Delta W$, the plus state $\mathcal{C}^{P}$ first encounters a supercritical pitchfork bifurcation.
At this point, the symmetry-preserving branch loses stability continuously and the system evolves onto an asymmetric branch $\mathcal{C}^{A}$.
Importantly, this symmetry-breaking bifurcation does not immediately trigger snap-through, as the asymmetric branch remains stable and allows the ribbon to evolve continuously under further transverse displacement.
The asymmetric branch subsequently reaches a saddle-node bifurcation, where it loses stability and triggers an abrupt transition toward the competing minus state $\mathcal{C}^{M}$.
Thus, symmetry breaking and snap-through occur at two distinct bifurcations, with the supercritical pitchfork initiating asymmetric deformation and the subsequent saddle-node triggering the snap-through event.
Representative configurations associated with the relevant equilibrium branches are shown in Fig.~\ref{fig:snap_16}(b).}
The corresponding dynamic response is presented in Fig.~\ref{fig:snap_16}(c).
{The normalized midpoint height $\Delta Z$ initially evolves smoothly as the ribbon departs from the symmetric configuration and follows the stable asymmetric branch, before undergoing a rapid change at snap-through.
The asymmetry measure $\mathcal{A}$ develops a pronounced peak near the snap-through event, indicating the involvement of asymmetric deformation in the dynamic transition.}
Representative experimental snapshots are shown in Fig.~\ref{fig:snap_16}(d).
{Together, the static bifurcation structure and dynamic response identify a two-stage instability pathway in which continuous symmetry breaking is followed by saddle-node-induced asymmetric top-to-bottom snap-through.}


\begin{figure}[!b]
    \centering
    \includegraphics[width=\textwidth]{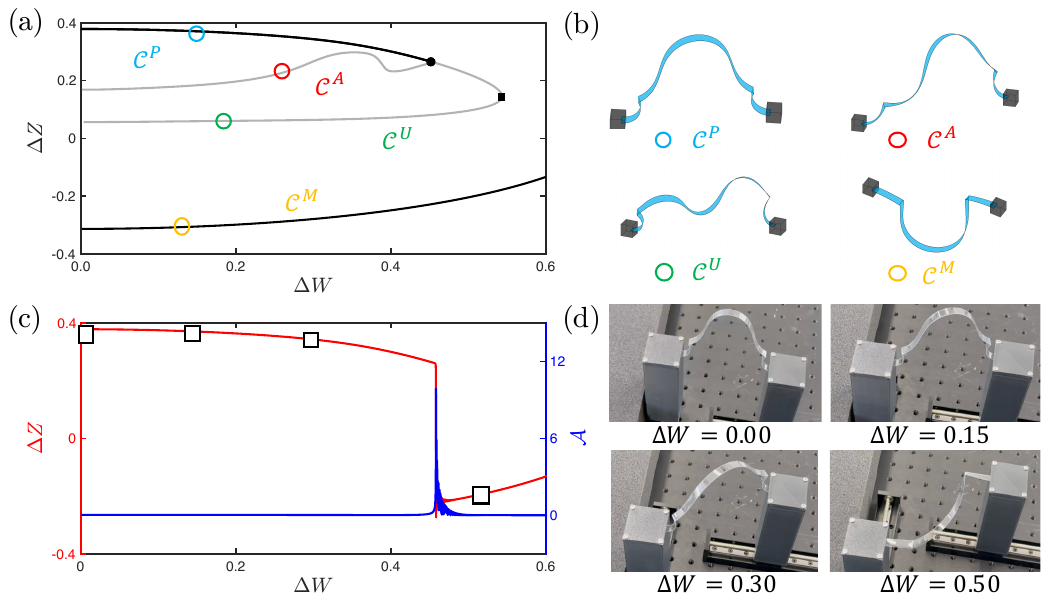}
    \caption{
    Subcritical pitchfork bifurcation preceding a saddle-node bifurcation and triggering asymmetric top-to-bottom snap-through for fold position $\Gamma=28/200=0.14$.
    (a) Static bifurcation diagram showing the normalized midpoint height $\Delta Z$ as a function of transverse displacement $\Delta W$.
    The plus state $\mathcal{C}^{P}$ first loses stability through a subcritical pitchfork bifurcation before reaching the saddle-node bifurcation, resulting in an abrupt asymmetric snap-through.
    (b) Representative numerical configurations associated with the plus state $\mathcal{C}^{P}$, asymmetric state $\mathcal{C}^{A}$, unstable state $\mathcal{C}^{U}$, and minus state $\mathcal{C}^{M}$.
    {The colored circles in (a) mark the corresponding equilibrium configurations shown in (b).}
    (c) Dynamic evolution of $\Delta Z$ and the asymmetry measure $\mathcal{A}$.
    The sharp growth of $\mathcal{A}$ during snap-through indicates the development of an asymmetric deformation.
    {The open squares mark the selected stages corresponding to the experimental snapshots in (d).}
    (d) Representative experimental snapshots during the transition.
    }
    \label{fig:snap_28}
\end{figure}

\paragraph{Subcritical-pitchfork-induced asymmetric snap-through} 
{A different asymmetric pathway is observed for $\Gamma=28/200=0.14$, as shown in Fig.~\ref{fig:snap_28}.
The static bifurcation diagram in Fig.~\ref{fig:snap_28}(a) reveals that the plus state $\mathcal{C}^{P}$ first encounters a subcritical pitchfork bifurcation, which occurs before the saddle-node bifurcation along the symmetric equilibrium branch.
Unlike the supercritical case, no nearby stable asymmetric continuation is available immediately beyond the bifurcation.
The system therefore departs abruptly from the symmetry-preserving branch and undergoes snap-through toward the minus state $\mathcal{C}^{M}$ through an asymmetric transient pathway.
In contrast to the preceding case, the first symmetry-breaking bifurcation itself triggers the loss of stability and the ensuing snap-through, without an intermediate stage of stable asymmetric deformation.}
Representative equilibrium configurations associated with $\mathcal{C}^{P}$, $\mathcal{C}^{A}$, $\mathcal{C}^{U}$, and $\mathcal{C}^{M}$ are shown in Fig.~\ref{fig:snap_28}(b).
{The dynamic simulation in Fig.~\ref{fig:snap_28}(c) confirms this interpretation.
The sudden change in $\Delta Z$ is accompanied by a sharp increase in $\mathcal{A}$, demonstrating that symmetry breaking occurs during the snap-through event.}
Representative experimental snapshots are shown in Fig.~\ref{fig:snap_28}(d).
{This pathway is therefore distinguished from the supercritical case by the direct onset of asymmetric snap-through at the subcritical pitchfork bifurcation, rather than at a subsequent saddle-node.}

\paragraph{Saddle-node-induced symmetric snap-through} 
{The third top-to-bottom pathway occurs for $\Gamma=42/200=0.21$, as shown in Fig.~\ref{fig:snap_42}.
The static bifurcation diagram in Fig.~\ref{fig:snap_42}(a) shows that the ordering of the competing bifurcations is reversed relative to the preceding case: the plus state $\mathcal{C}^{P}$ reaches a saddle-node bifurcation before the subcritical pitchfork bifurcation.
The stable symmetric branch therefore terminates while the ribbon remains symmetry-preserving.
Beyond this limit point, the system can no longer remain on the upper equilibrium branch and snaps toward the minus state $\mathcal{C}^{M}$.
Although asymmetric equilibrium branches exist elsewhere in the bifurcation structure, they are not encountered before the primary saddle-node instability.
The dynamic response in Fig.~\ref{fig:snap_42}(c) shows a sudden change in $\Delta Z$, while the asymmetry measure $\mathcal{A}$ remains approximately zero throughout the transition.
This confirms that the snap-through proceeds through a symmetry-preserving pathway.
The corresponding representative configurations are shown in Fig.~\ref{fig:snap_42}(b), while experimental snapshots are presented in Fig.~\ref{fig:snap_42}(d).
The comparison with the subcritical-pitchfork-induced pathway demonstrates that reversing the order of the pitchfork and saddle-node bifurcations changes the snap-through response from asymmetric to symmetric.

\begin{figure}[h]
    \centering
    \includegraphics[width=\textwidth]{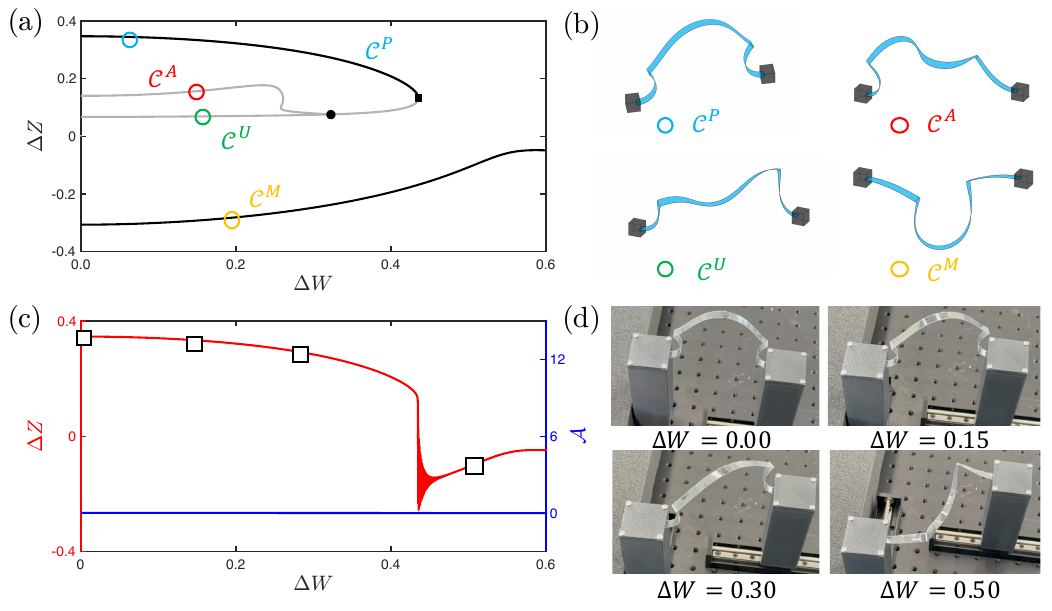}
    \caption{
    Saddle-node bifurcation preceding a subcritical pitchfork bifurcation and triggering symmetric top-to-bottom snap-through for fold position $\Gamma=42/200=0.21$.
    (a) Static bifurcation diagram showing the normalized midpoint height $\Delta Z$ as a function of transverse displacement $\Delta W$.
    The plus state $\mathcal{C}^{P}$ first loses stability at a saddle-node bifurcation before reaching the subcritical pitchfork bifurcation, resulting in symmetric snap-through.
    (b) Representative numerical configurations associated with the plus state $\mathcal{C}^{P}$, asymmetric state $\mathcal{C}^{A}$, unstable state $\mathcal{C}^{U}$, and minus state $\mathcal{C}^{M}$.
    {The colored circles in (a) mark the corresponding equilibrium configurations shown in (b).}
    (c) Dynamic evolution of $\Delta Z$ and the asymmetry measure $\mathcal{A}$.
    The near-zero value of $\mathcal{A}$ throughout the transition indicates that the snap-through remains symmetric.
    {The open squares mark the selected stages corresponding to the experimental snapshots in (d).}
    (d) Representative experimental snapshots during the transition.
    }
    \label{fig:snap_42}
\end{figure}

The robustness of the asymmetric and symmetric snap-through
pathways against variations in damping, loading speed, initial
imperfections, and folding stiffness is further examined in~\ref{sec:AppendixParameter}.}

\subsection{Fold-position-induced transitions between snap-through regimes}

\begin{figure}[h]
    \centering
    \includegraphics[width=\textwidth]{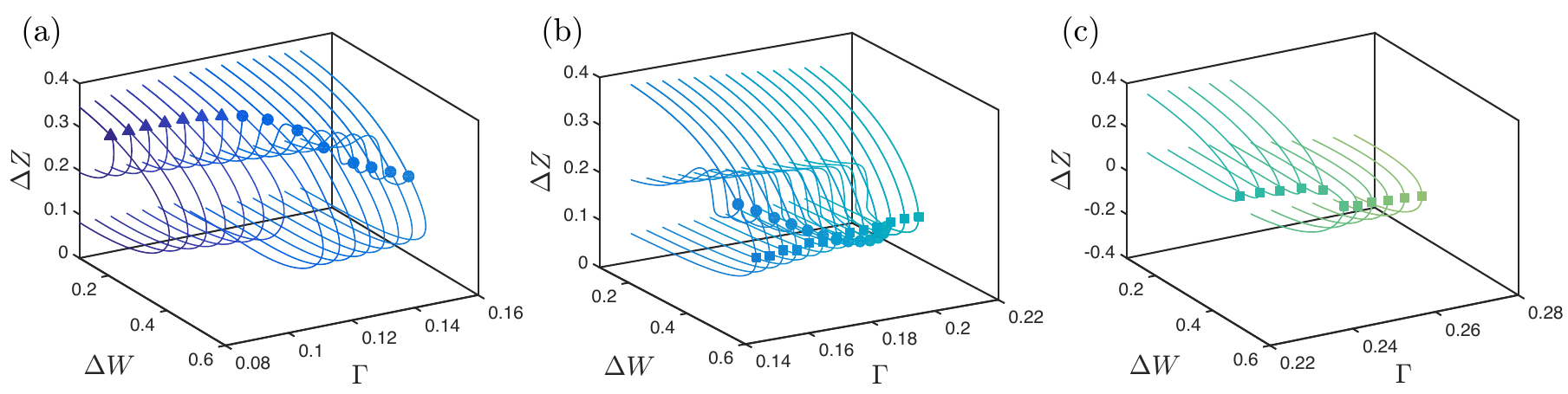}
    \caption{
    {Fold-position-induced transitions between distinct snap-through regimes.
    Three-dimensional equilibrium bifurcation diagrams showing the evolution of the normalized midpoint height $\Delta Z$ with transverse displacement $\Delta W$ and fold position $\Gamma$.
    (a) Transition from Regime I to Regime II, where the first pitchfork bifurcation changes from supercritical to subcritical, replacing continuous symmetry breaking followed by saddle-node-induced snap-through with pitchfork-triggered asymmetric snap-through.
    (b) Transition from Regime II to Regime III, where the ordering of the subcritical pitchfork and saddle-node bifurcations reverses, switching the top-to-bottom snap-through pathway from asymmetric to symmetric.
    (c) Transition from Regime III to Regime IV, where the destabilized metastable state switches from the plus state $\mathcal{C}^{P}$ to the minus state $\mathcal{C}^{M}$, reversing the snap-through direction while retaining a saddle-node-induced symmetric pathway.}
    }
    \label{fig:global_transition}
\end{figure}

{The three top-to-bottom snap-through pathways identified above represent distinct regimes of a continuously evolving bifurcation landscape controlled by fold position $\Gamma$.
We denote these pathways as Regimes I--III, characterized by a supercritical pitchfork followed by saddle-node-induced asymmetric snap-through, a subcritical-pitchfork-induced asymmetric transition, and a saddle-node-induced symmetric transition, respectively.
To understand how these regimes are connected, we track the evolution of the equilibrium branches and their stability as $\Gamma$ varies, as shown in Fig.~\ref{fig:global_transition}.
Three distinct transitions emerge, involving changes in pitchfork criticality, bifurcation ordering, and the identity of the destabilized metastable state.}

Figure~\ref{fig:global_transition}(a) shows the transition from Regime I to Regime II.
{In Regime I, the supercritical pitchfork gives rise to stable asymmetric branches, allowing the ribbon to depart continuously from the symmetric configuration before reaching a subsequent saddle-node instability.
As $\Gamma$ increases, the pitchfork changes from supercritical to subcritical, and the asymmetric branches emerging from the bifurcation become locally unstable.
The symmetric state therefore loses stability without access to a nearby stable asymmetric continuation, directly triggering asymmetric snap-through in Regime II.
This change in pitchfork criticality alters the relationship between symmetry breaking and snap-through: they occur at separate bifurcations in Regime I but coincide at the first instability in Regime II.
Nevertheless, both regimes retain asymmetric top-to-bottom snap-through, despite their different instability pathways.}

{Figure~\ref{fig:global_transition}(b) illustrates the transition from Regime II to Regime III.
As $\Gamma$ increases, the critical transverse displacements associated with the subcritical pitchfork and saddle-node bifurcations evolve until their relative ordering reverses.
In Regime II, the pitchfork is encountered first, and the loss of stability of the symmetric branch initiates an asymmetric transition.
In Regime III, the saddle-node instead occurs first, terminating the stable symmetric branch before the symmetry-breaking bifurcation can be reached along the loading path.
The first instability thus changes from a symmetry-breaking pitchfork to a symmetry-preserving saddle-node, switching the top-to-bottom snap-through pathway from asymmetric to symmetric.
This transition demonstrates that the relative ordering of competing bifurcations determines which instability mechanism governs the observed snap-through response.
This bifurcation-reordering mechanism is further illustrated
by a reduced-order double-mass von Mises truss model, which
reproduces the transition between asymmetric and symmetric
snap-through pathways (see~\ref{sec:AppendixToymodel}).

Figure~\ref{fig:global_transition}(c) reveals a different transition between Regimes III and IV.
Unlike the preceding transitions, which modify the instability pathway of the plus state, this transition changes which of the two metastable states becomes unstable under transverse displacement.
In Regime III, the plus state $\mathcal{C}^{P}$ reaches a saddle-node instability and undergoes symmetric top-to-bottom snap-through.
As the fold position increases, the relevant saddle-node instability becomes associated with the minus state $\mathcal{C}^{M}$, giving rise to Regime IV, in which the ribbon snaps upward toward $\mathcal{C}^{P}$.
The transition therefore reverses the direction of state switching while preserving both the saddle-node instability mechanism and the symmetry of the snap-through pathway.
This illustrates an additional level of geometric control, whereby localized folds determine not only how instability occurs but also which metastable state loses stability.}

\subsection{Bottom-to-top snap-through pathways}

{As the fold position increases further, the stability landscape is reorganized such that the minus state $\mathcal{C}^{M}$ becomes the state that loses stability under increasing transverse displacement.
The snap-through direction is therefore reversed, and the ribbon transitions from $\mathcal{C}^{M}$ to $\mathcal{C}^{P}$.
As shown in Fig.~\ref{fig:bottom_to_top}, three distinct instability-mediated pathways also emerge for this reverse transition, corresponding to saddle-node-induced symmetric snapping, supercritical-pitchfork-first asymmetric snapping, and subcritical-pitchfork-induced asymmetric snapping.
These pathways are distinguished by the nature and sequence of the instabilities encountered along the initially stable minus branch.

\begin{figure}[!h]
    \centering
    \includegraphics[width=\textwidth]{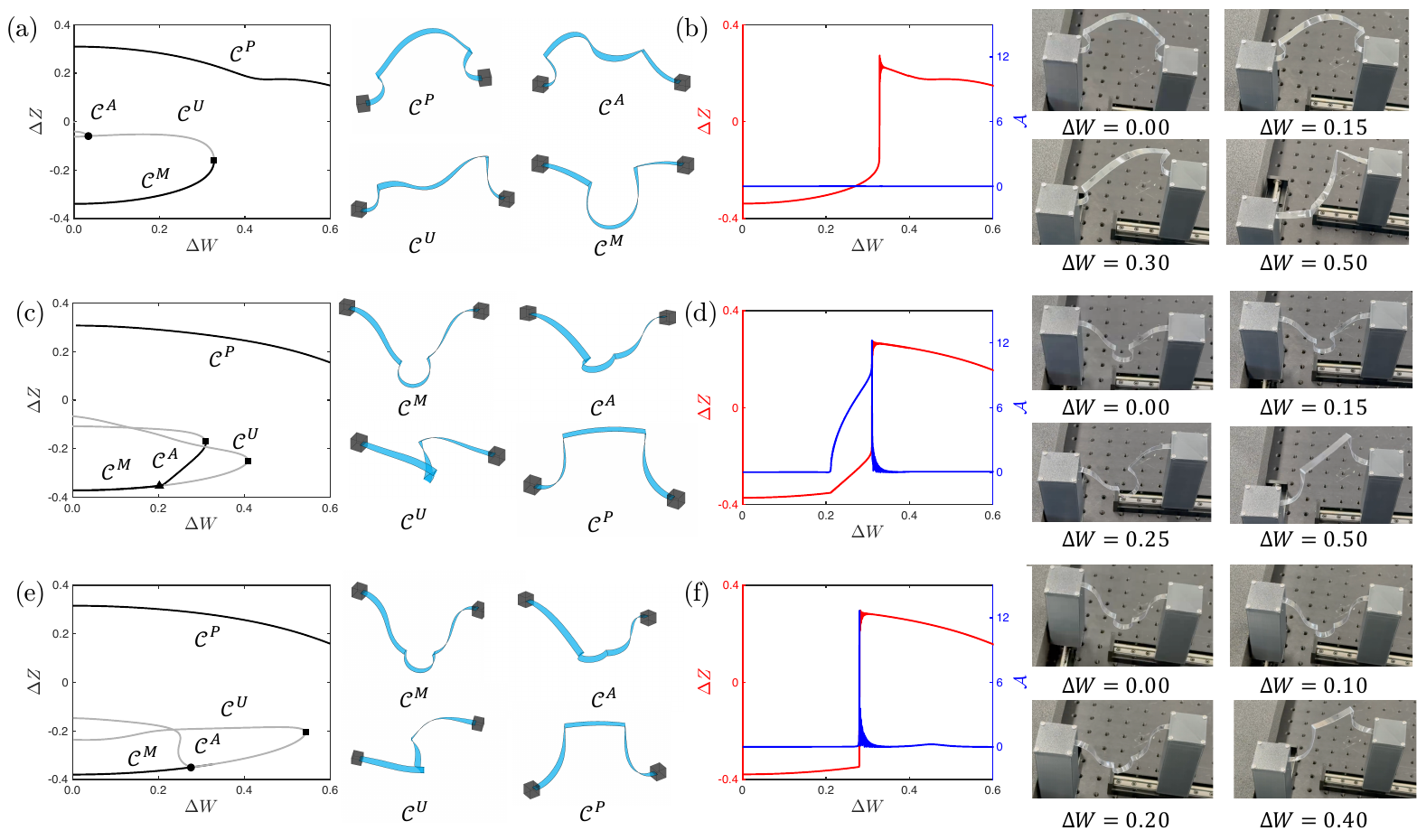}
    \caption{
    {Three distinct bottom-to-top snap-through regimes governed by different bifurcation sequences.
    (a, b) Regime IV: saddle-node-induced symmetric bottom-to-top snap-through for fold position $\Gamma=56/200=0.28$.
    (a) Static bifurcation diagram showing the normalized midpoint height $\Delta Z$ as a function of transverse displacement $\Delta W$, together with representative numerical configurations of the plus state $\mathcal{C}^{P}$, asymmetric state $\mathcal{C}^{A}$, unstable state $\mathcal{C}^{U}$, and minus state $\mathcal{C}^{M}$.
    The minus state $\mathcal{C}^{M}$ first loses stability at a saddle-node bifurcation, triggering symmetric bottom-to-top snap-through.
    (b) Dynamic evolution of $\Delta Z$ and the asymmetry measure $\mathcal{A}$, together with representative experimental snapshots.
    (c, d) Regime V: supercritical-pitchfork-first asymmetric bottom-to-top snap-through for fold position $\Gamma=68/200=0.34$.
    (c) Static bifurcation diagram and representative numerical configurations.
    The minus state $\mathcal{C}^{M}$ first undergoes a supercritical pitchfork bifurcation, followed by evolution along a stable asymmetric branch and subsequent saddle-node-induced snap-through.
    (d) Dynamic evolution of $\Delta Z$ and $\mathcal{A}$, together with representative experimental snapshots.
    (e, f) Regime VI: subcritical-pitchfork-first asymmetric bottom-to-top snap-through for fold position $\Gamma=74/200=0.37$.
    (e) Static bifurcation diagram and representative numerical configurations.
    The minus state $\mathcal{C}^{M}$ first loses stability through a subcritical pitchfork bifurcation, triggering asymmetric bottom-to-top snap-through.
    (f) Dynamic evolution of $\Delta Z$ and $\mathcal{A}$, together with representative experimental snapshots.}
    }
    \label{fig:bottom_to_top}
\end{figure}

The reversal of snap-through direction is first illustrated by Regime IV at $\Gamma=56/200=0.28$, where the minus state $\mathcal{C}^{M}$ reaches a saddle-node bifurcation before any symmetry-breaking instability, as shown in Fig.~\ref{fig:bottom_to_top}(a).
The stable symmetric branch terminates at this limit point, leaving no nearby stable continuation and triggering snap-through toward the plus state $\mathcal{C}^{P}$.
Because the saddle-node occurs before the pitchfork bifurcation, the ribbon remains symmetric up to the onset of instability.
The dynamic response and experimental snapshots in Fig.~\ref{fig:bottom_to_top}(b) show an abrupt reversal of $\Delta Z$, while the asymmetry measure $\mathcal{A}$ remains close to zero throughout the transition, confirming a symmetry-preserving bottom-to-top snap-through.
This regime is the reverse counterpart of the saddle-node-induced symmetric top-to-bottom snapping observed in Regime III, with the same instability mechanism but a different destabilized metastable state.

A qualitatively different pathway emerges in Regime V ($\Gamma=68/200=0.34$), where a supercritical pitchfork bifurcation precedes the bottom-to-top snap-through, as shown in Fig.~\ref{fig:bottom_to_top}(c).
The minus state initially loses symmetry continuously and evolves along a stable asymmetric branch under further transverse displacement.
This asymmetric deformation remains stable until a subsequent saddle-node bifurcation is reached, at which point the ribbon undergoes an abrupt bottom-to-top snap-through.
The dynamic evolution in Fig.~\ref{fig:bottom_to_top}(d) shows a gradual increase in the asymmetry measure $\mathcal{A}$ before the rapid change in $\Delta Z$, consistent with the two-stage instability pathway identified in Regime I.
Thus, the onset of symmetry breaking does not necessarily coincide with the onset of snap-through, even when the transition direction is reversed.}

{Further increasing the fold position to $\Gamma=74/200=0.37$ leads to Regime VI, in which the first instability becomes a subcritical pitchfork bifurcation, as shown in Fig.~\ref{fig:bottom_to_top}(e).
In contrast to Regime V, the asymmetric branches emerging from the pitchfork are locally unstable, so the ribbon cannot evolve continuously onto a nearby stable asymmetric configuration.
The first symmetry-breaking instability therefore directly triggers asymmetric bottom-to-top snap-through, rather than allowing a separate stage of stable asymmetric deformation.
As shown in Fig.~\ref{fig:bottom_to_top}(f), the abrupt change in $\Delta Z$ is accompanied by a sharp increase in $\mathcal{A}$, demonstrating that symmetry breaking and snap-through occur together, as in Regime II.

Taken together, these results establish that the three instability-mediated pathways identified for top-to-bottom snapping also arise in the reverse direction, although their ordering with increasing fold position differs.
The correspondence between Regimes I and V, II and VI, and III and IV highlights the distinction between the mechanism of instability and the direction of state switching.
Fold placement therefore provides two complementary levels of control: selecting which metastable state becomes unstable and determining whether the resulting transition is governed by a supercritical pitchfork followed by a saddle-node, a subcritical pitchfork, or a saddle-node alone.
Supplementary videos of the six representative snap-through
pathways are provided in~\ref{sec:AppendixVideo}.}

\subsection{Global organization of snap-through regimes}

\begin{figure}[!t]
    \centering
    \includegraphics[width=0.9\textwidth]{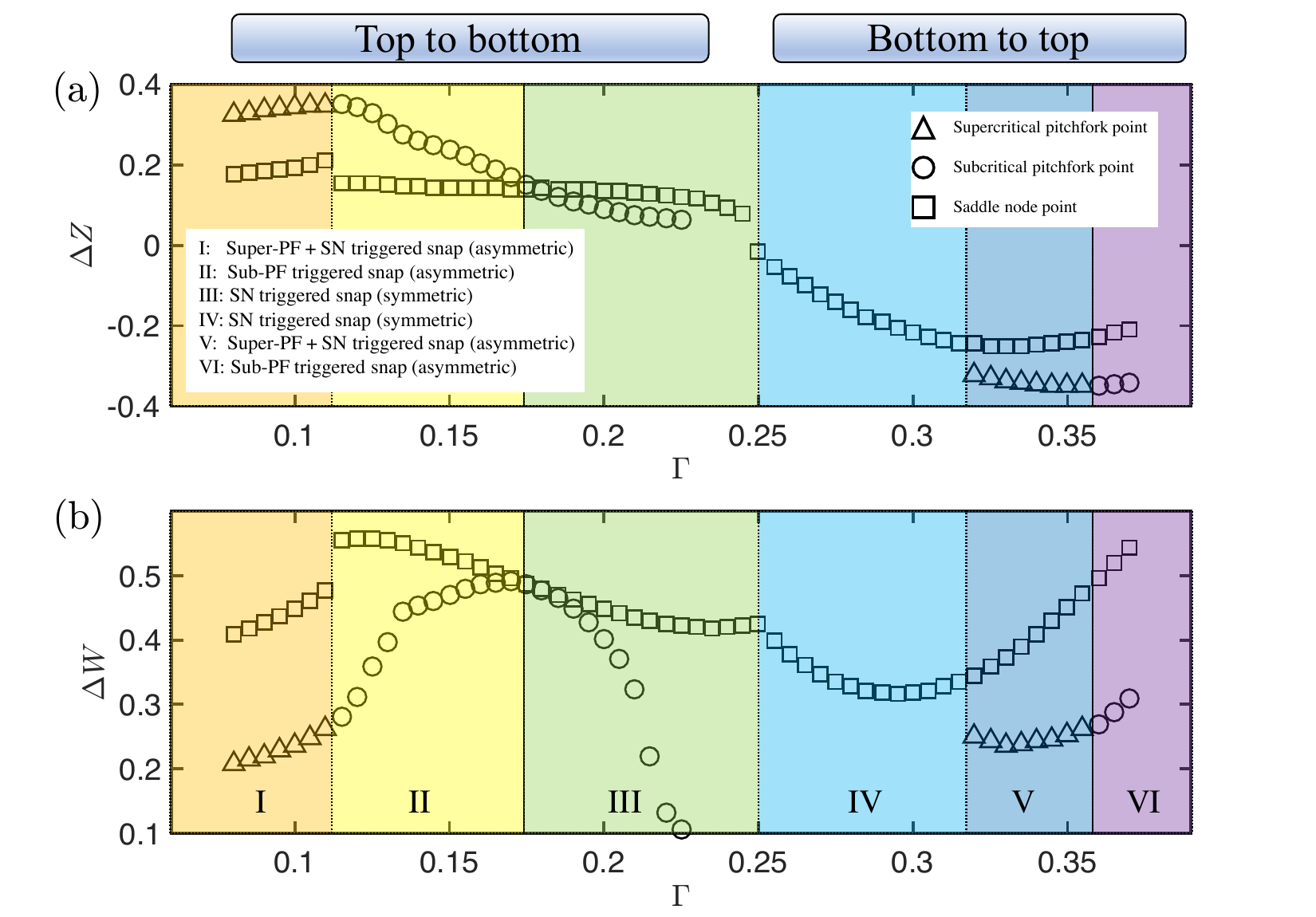}
    \caption{
    {Global organization of six snap-through regimes controlled by fold position $\Gamma$.
    (a) Critical midpoint height $\Delta Z$ as a function of $\Gamma$, showing the evolution of supercritical pitchfork (triangles), subcritical pitchfork (circles), and saddle-node (squares) bifurcation points.
    (b) Corresponding critical transverse displacement $\Delta W$ as a function of $\Gamma$, showing how the bifurcation thresholds vary with fold position.
    The six shaded regions (I--VI) represent distinct snap-through regimes, classified according to the bifurcation sequence, snap-through symmetry, and transition direction.
    Regimes I--III correspond to top-to-bottom snap-through, whereas Regimes IV--VI correspond to bottom-to-top snap-through.
    The boundaries between adjacent regimes mark changes in pitchfork criticality (I--II and V--VI), bifurcation ordering (II--III and IV--V), and the destabilized metastable state (III--IV).
    The markers represent selected characteristic bifurcation points associated with the snap-through pathways and regime transitions.}
    }
    \label{fig:global_map}
\end{figure}

{The preceding results reveal six distinct snap-through regimes arising from the geometric reorganization of competing instabilities.
To establish their global organization, we systematically track the critical bifurcation points over a broad range of fold positions $\Gamma$, as summarized in Fig.~\ref{fig:global_map}.
The two complementary maps describe the evolution of the critical midpoint configurations and transverse displacement thresholds, providing a unified picture of how fold placement controls the snap-through response.}

{Figure~\ref{fig:global_map}(a) presents the normalized midpoint height $\Delta Z$ at the critical bifurcation points as a function of $\Gamma$.
The supercritical pitchfork, subcritical pitchfork, and saddle-node bifurcations are distinguished by triangular, circular, and square markers, respectively.
The critical configurations evolve continuously with fold position, while the bifurcation types and their associated equilibrium branches change across the different regimes.
For smaller $\Gamma$, the relevant snap-through instability occurs on the positive-$\Delta Z$ branch, corresponding to top-to-bottom transitions in Regimes I--III.
Beyond the III--IV boundary, the destabilized state switches to the negative-$\Delta Z$ branch, giving rise to bottom-to-top transitions in Regimes IV--VI.
This change in the critical configuration distinguishes the reversal of snap-through direction from the changes in bifurcation mechanism occurring within either transition direction.

Figure~\ref{fig:global_map}(b) shows the critical transverse displacement $\Delta W$ as a function of $\Gamma$.
Multiple bifurcation thresholds may coexist at a given fold position, reflecting the presence of competing instability mechanisms within the same equilibrium landscape.
The observed snap-through pathway is therefore determined not simply by the existence of a particular bifurcation, but by which instability is encountered first along the initially stable loading path and whether a stable continuation remains accessible beyond it.
For example, in Regimes II and III, the relative ordering of the subcritical pitchfork and saddle-node thresholds determines whether symmetry breaking or a symmetry-preserving instability governs snap-through.
In Regimes I and V, the supercritical pitchfork instead permits continuous evolution along a stable asymmetric branch before a subsequent saddle-node triggers snap-through.
The critical-displacement map thus complements the critical-configuration map by revealing how competing thresholds select different instability pathways.

The six-regime structure can therefore be interpreted in terms of three distinct characteristics of the stability landscape.
First, the \emph{identity of the destabilized branch} determines the direction of snap-through: destabilization of $\mathcal{C}^{P}$ gives a top-to-bottom transition, whereas destabilization of $\mathcal{C}^{M}$ gives a bottom-to-top transition.
Second, the \emph{type and ordering of the first instability} determine the symmetry of the transition: a saddle-node-first instability produces a symmetry-preserving pathway, whereas a pitchfork-first instability produces symmetry breaking before or during snap-through.
Third, within the pitchfork-controlled regimes, the \emph{criticality of the pitchfork bifurcation} distinguishes a continuous development of asymmetry from an abrupt symmetry-breaking departure.
Together, these characteristics yield a unified $2\times3$ classification, combining two snap-through directions with three instability-mediated pathways.
The experimentally observed pathways at six representative fold positions are consistent with this classification, while the regime boundaries are determined through numerical continuation.
This classification demonstrates that localized fold placement does more than shift individual instability thresholds: it reorganizes the bifurcation landscape to control when snap-through occurs, how the transition proceeds, and which metastable state is destabilized.}

\section{Conclusions}
\label{sec:discussion}

{In this work, we investigated transverse-displacement-induced snap-through in folded elastic ribbons and showed that localized fold placement can reorganize the bifurcation landscape to control both the symmetry and direction of the transition.
By combining experiments, discrete models, dynamic simulations, and numerical continuation, we identified six snap-through regimes arising from two transition directions---top-to-bottom and bottom-to-top---and three characteristic instability pathways: a supercritical pitchfork followed by stable asymmetric deformation and subsequent saddle-node-induced snap-through, a subcritical pitchfork directly triggering asymmetric snap-through, and a saddle-node inducing symmetric snap-through.

We further showed that varying the fold position changes the pitchfork criticality, the relative ordering of pitchfork and saddle-node bifurcations, and which metastable branch loses stability.
These three mechanisms govern different aspects of the snap-through response: the relationship between symmetry breaking and snap-through, the symmetry of the transition pathway, and the direction of state switching, respectively.
The resulting phase diagram provides a unified picture of these transitions and demonstrates that localized folds can program not only how the ribbon snaps, but also which state undergoes the transition.
These findings provide a geometric strategy for controlling state transitions in multistable and morphing structures.}

\section*{CRediT authorship contribution statement}
\textbf{Weicheng Huang}: Writing – Original draft, Methodology, Software, Formal analysis, Investigation, Data curation, Validation, Project administration, Funding acquisition, Conceptualization. 
\textbf{Qun Zhang}: Formal analysis, Investigation, Validation.  
\textbf{Bohan Zhang}: Writing – Review \& Editing, Formal analysis, Investigation, Validation. 
\textbf{Mingchao Liu}: Writing – Review \& Editing, Methodology, Formal analysis, Investigation, Validation, Supervision, Project administration, Funding acquisition, Conceptualization.

\section*{Declaration of competing interest}
The authors declare that they have no known competing financial interests or personal relationships that could have appeared to influence the work reported in this paper.

\section*{Acknowledgments}

W.H. acknowledges the start-up funding from Newcastle University, UK. M.L. acknowledges start-up funding from The University of Birmingham, UK.

\appendix

\counterwithin{figure}{section}
\renewcommand{\thefigure}{\Alph{section}.\arabic{figure}}

\section{Convergence study}
\label{sec:AppendixConvergent}

\begin{figure}[!h]
    \centering
    \includegraphics[width=\textwidth]{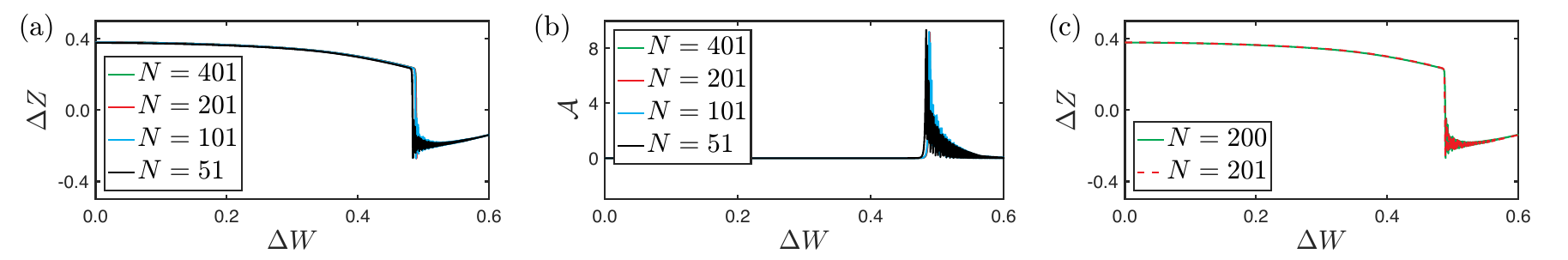}
    \caption{
    {Mesh-convergence study for the folded ribbon with $\Gamma=0.14$.
    (a,b) Dynamic evolution of (a) the normalized midpoint height $\Delta Z$ and (b) the asymmetry measure $\mathcal{A}$ as functions of transverse displacement $\Delta W$, for discretizations with $N=51$, $101$, $201$, and $401$ nodes.
    The results show convergence of the snap-through threshold and the overall dynamic response with increasing mesh resolution.
    (c) Comparison between even and odd discretizations ($N=200$ and $N=201$), showing nearly identical midpoint responses and confirming that the snap-through behavior is insensitive to the parity of $N$.}
    }
    \label{fig:convergence}
\end{figure}

{A mesh-convergence study was performed for a representative case with fold position $\Gamma=0.14$, viscosity coefficient $\mu=0.1$, imperfection amplitude $\delta \alpha = 0.001$, and transverse loading speed $\delta \dot{w}  =  0.01$ m/s.
As shown in Fig.~\ref{fig:convergence}(a,b), the dynamic responses were compared using $N=51$, $101$, $201$, and $401$ nodes.
The predicted evolution of the normalized midpoint displacement $\Delta Z$, the onset of snap-through, and the growth of the asymmetry measure $\mathcal{A}$ show good convergence with increasing spatial resolution.
In particular, the results for $N=101$, $201$, and $401$ are nearly indistinguishable before and through the instability, while the coarser discretization with $N=51$ mainly produces larger post-snap oscillations without changing the predicted transition point or snap-through pathway.
We further compared an even and odd number of nodes, $N=200$ and $N=201$, as shown in Fig.~\ref{fig:convergence}(c).
The two responses are essentially identical, demonstrating that the numerical prediction is insensitive to whether the discretization contains an even or odd number of nodes.
Based on these results, $N=201$ is adopted throughout the simulations as a suitable compromise between numerical accuracy and computational efficiency.}

\section{Parameter sensitivity study}
\label{sec:AppendixParameter}

\begin{figure}[!b]
    \centering
    \includegraphics[width=0.85\textwidth]{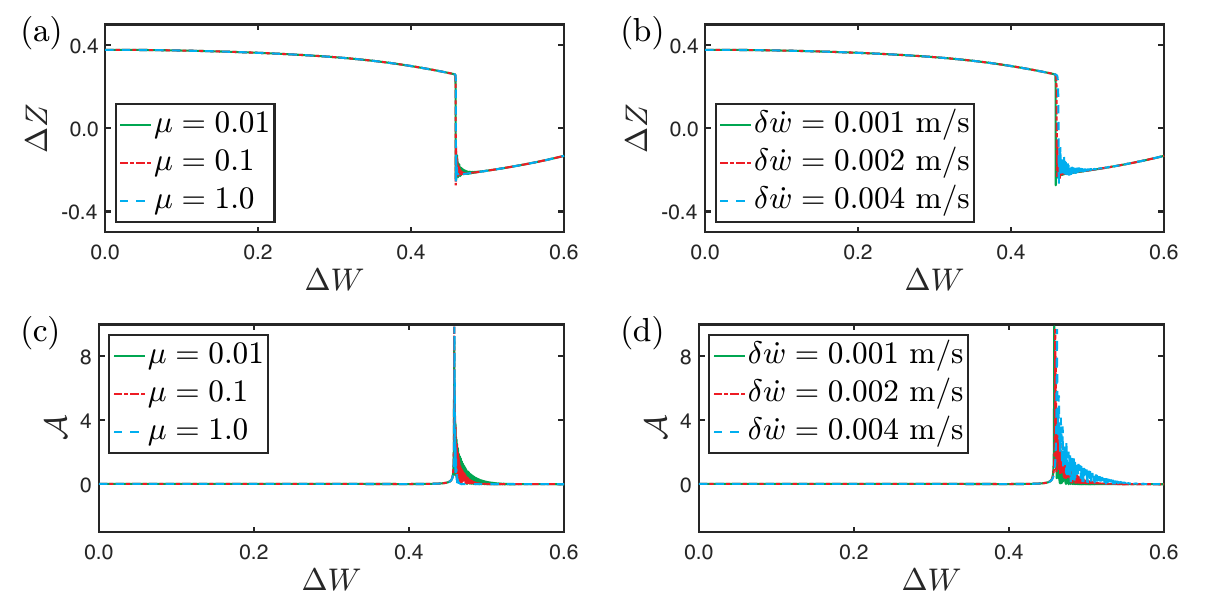}
    \caption{
    {Sensitivity of the dynamic snap-through response to damping and loading speed for the asymmetric case with fold position $\Gamma=0.14$.
    (a,c) Effect of the mass-proportional damping coefficient $\mu$ on (a) the normalized midpoint height $\Delta Z$ and (c) the asymmetry measure $\mathcal{A}$.
    (b,d) Corresponding effect of the transverse loading speed $\delta\dot{w}$ on (b) $\Delta Z$ and (d) $\mathcal{A}$.}
    }
    \label{fig:parameter_a1}
\end{figure}

\begin{figure}[!b]
    \centering
    \includegraphics[width=0.85\textwidth]{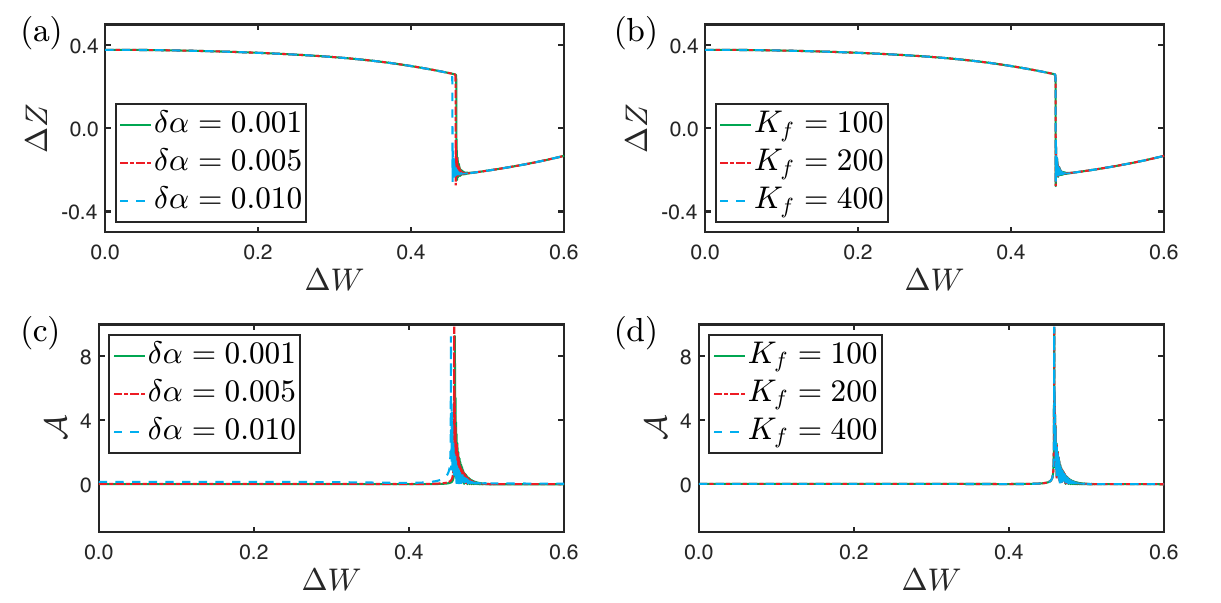}
    \caption{
    {Sensitivity of the dynamic snap-through response to initial imperfections and folding stiffness for the asymmetric case with fold position $\Gamma=0.14$.
    (a,c) Effect of the initial imperfection amplitude $\delta\alpha$ on (a) the normalized midpoint height $\Delta Z$ and (c) the asymmetry measure $\mathcal{A}$.
    (b,d) Corresponding effect of the folding stiffness $K_f$ on (b) $\Delta Z$ and (d) $\mathcal{A}$.}
    }
    \label{fig:parameter_a2}
\end{figure}

\begin{figure}[!h]
    \centering
    \includegraphics[width=0.85\textwidth]{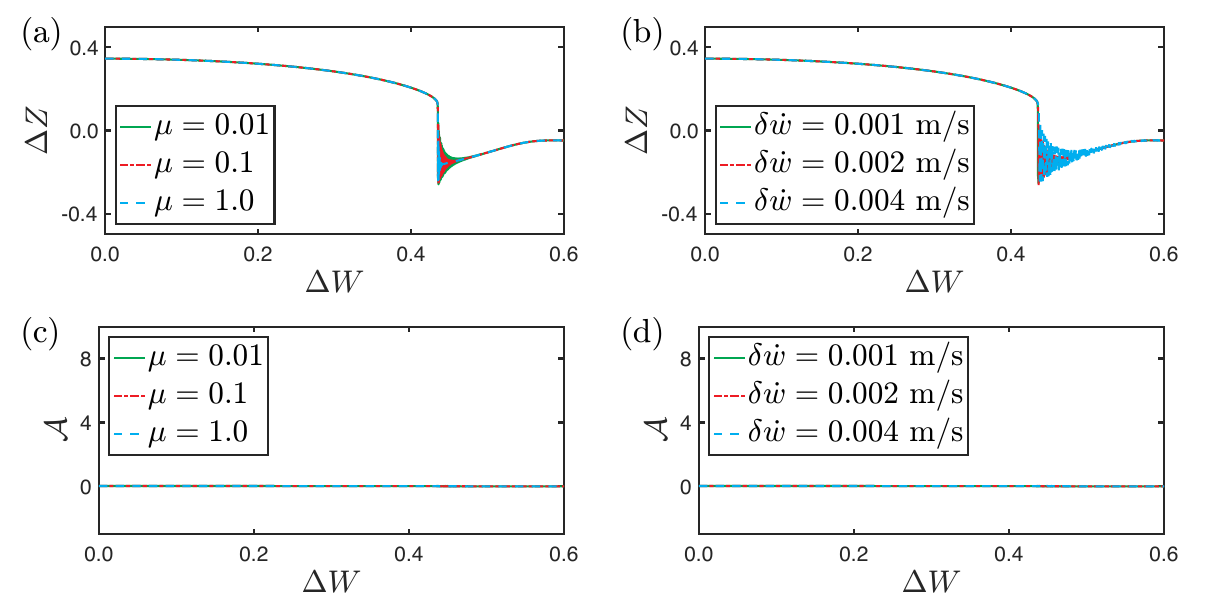}
    \caption{
    {Sensitivity of the dynamic snap-through response to damping and loading speed for the symmetric case with fold position $\Gamma=0.21$.
    (a,c) Effect of the mass-proportional damping coefficient $\mu$ on (a) the normalized midpoint height $\Delta Z$ and (c) the asymmetry measure $\mathcal{A}$.
    (b,d) Corresponding effect of the transverse loading speed $\delta\dot{w}$ on (b) $\Delta Z$ and (d) $\mathcal{A}$.}
    }
    \label{fig:parameter_s1}
\end{figure}

\begin{figure}[!h]
    \centering
    \includegraphics[width=0.85\textwidth]{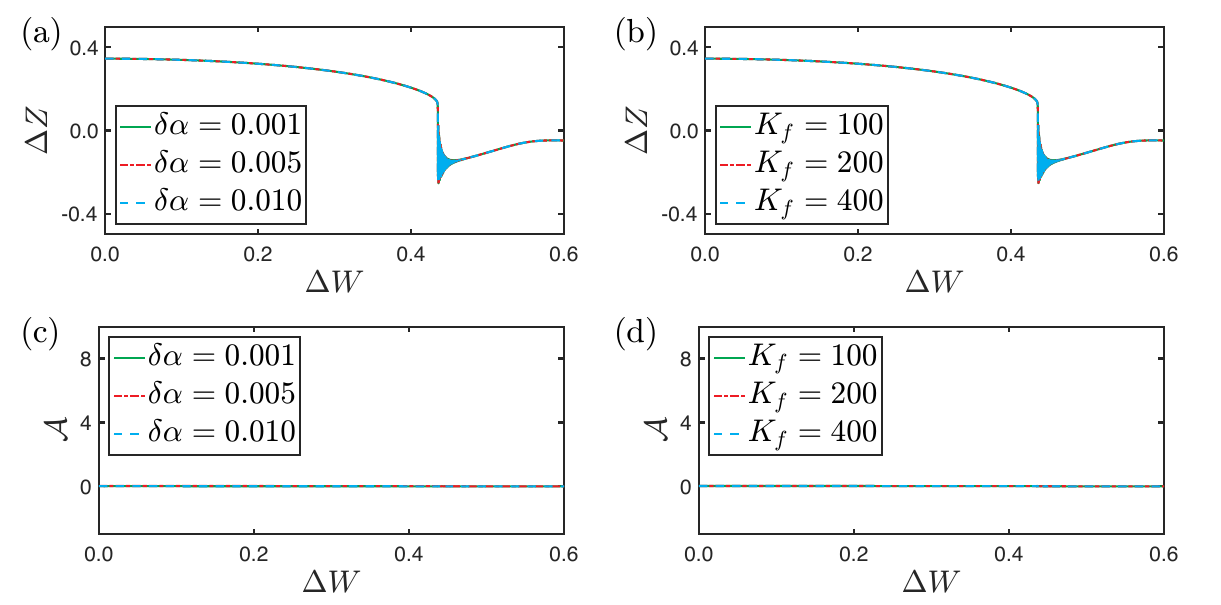}
    \caption{
    {Sensitivity of the dynamic snap-through response to initial imperfections and folding stiffness for the symmetric case with fold position $\Gamma=0.21$.
    (a,c) Effect of the initial imperfection amplitude $\delta\alpha$ on (a) the normalized midpoint height $\Delta Z$ and (c) the asymmetry measure $\mathcal{A}$.
    (b,d) Corresponding effect of the folding stiffness $K_f$ on (b) $\Delta Z$ and (d) $\mathcal{A}$.}
    }
    \label{fig:parameter_s2}
\end{figure}

{To examine the robustness of the predicted snap-through pathways, we performed a parameter sensitivity study considering the effects of viscous damping, loading speed, initial imperfection, and folding stiffness, as shown in Figs.~\ref{fig:parameter_a1}--\ref{fig:parameter_s2}.
Two representative cases were considered: an asymmetric snap-through case with $\Gamma=0.14$ and a symmetric snap-through case with $\Gamma=0.21$.}

{%
For the asymmetric case, variations in the damping coefficient and loading speed have little influence on the overall evolution of the midpoint height $\Delta Z$ and the snap-through threshold, but affect the magnitude and relaxation of the transient asymmetry response $\mathcal{A}$.
Changes in the initial imperfection amplitude and folding stiffness similarly leave the overall snap-through response largely unchanged, although small differences are observed in the transient asymmetry, as shown in Figs.~\ref{fig:parameter_a1}--\ref{fig:parameter_a2}.
In all cases, $\mathcal{A}$ exhibits a pronounced increase during snap-through, demonstrating the persistence of the asymmetric transition pathway.
For the symmetric case, the midpoint response remains nearly identical across the tested parameter ranges, while $\mathcal{A}$ stays close to zero throughout the transition, as shown in Figs.~\ref{fig:parameter_s1}--\ref{fig:parameter_s2}.}

{%
These results indicate that although the numerical parameters can influence the quantitative transient dynamics, the distinction between symmetric and asymmetric snap-through pathways remains robust over the parameter ranges examined.
This supports the conclusion that the observed pathway selection is primarily governed by the underlying bifurcation structure rather than the particular choices of damping, loading speed, initial imperfection, or folding stiffness.}

\section{Reduced-order model}
\label{sec:AppendixToymodel}

To rationalize the bifurcation reordering mechanism observed in the folded ribbon, we adopt a reduced-order double-mass von Mises truss model \citep{wang2024transient,giudici2025transient}. The model is not intended to reproduce the full geometry of the ribbon, but to capture the minimal competition between a symmetry-breaking pitchfork bifurcation and a symmetry-preserving saddle-node bifurcation.

\begin{figure}[!h]
    \centering
    \includegraphics[width=\textwidth]{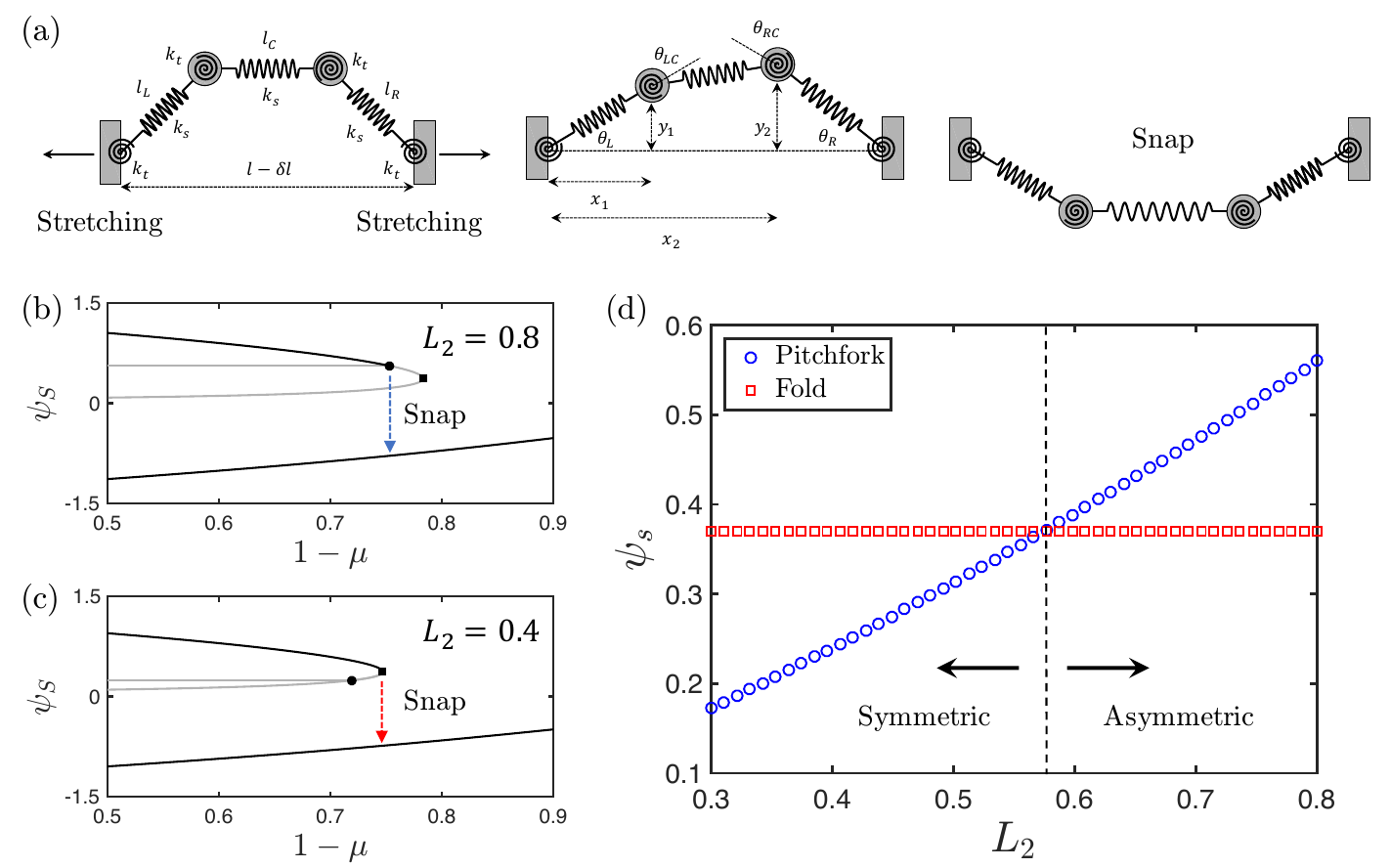}
    \caption{
    Reduced-order double-mass von Mises truss model illustrating bifurcation reordering and snap-through symmetry selection.
    (a) Schematic of the model, consisting of two internal masses connected by linear and torsional springs. The geometric parameter $L_2$ controls the equilibrium bifurcation structure and the competition between symmetry-breaking and symmetry-preserving instabilities.
    (b,c) Static bifurcation diagrams for (b) $L_2=0.8$ and (c) $L_2=0.4$, showing asymmetric and symmetric snap-through pathways, respectively. Black and grey curves denote stable and unstable equilibrium branches, while circular and square markers indicate pitchfork and fold (saddle-node) bifurcations, respectively. Changing $L_2$ reverses the ordering of the competing instabilities, thereby switching the snap-through pathway.
    (d) Critical values of $\psi_s$ associated with the pitchfork and fold bifurcations as functions of $L_2$. The dashed line marks the crossing of the two critical thresholds, separating the saddle-node-first symmetric regime from the pitchfork-first asymmetric regime.
    }
    \label{fig:toy_model}
\end{figure}

As shown in Fig.~\ref{fig:toy_model}(a), the system consists of two point masses connected to the left and right supports by two linear springs of stiffness $k_s$ and natural length $l_1$. The instantaneous lengths of these two side springs are denoted by $l_L$ and $l_R$, respectively. The two masses are further connected by a central spring with the same stiffness $k_s$ and natural length $l_2$, whose instantaneous length is denoted by $l_C$. Torsional springs of stiffness $k_t$ are placed at the two supports and at the two internal joints. The support torsional springs have a prescribed rest angle $\alpha$, whereas the internal torsional springs are relaxed when the adjacent springs are aligned. The geometric parameter $l_2$ controls the relative spacing between the two masses and plays a role analogous to the fold position in the ribbon system.
Starting from the upper equilibrium branch, we gradually release the imposed compression and track the equilibrium paths and stability changes. This loading protocol allows us to determine whether the pitchfork or saddle-node bifurcation is encountered first, and hence whether the model predicts an asymmetric or symmetric snap-through pathway.

\paragraph{Kinematics and energy formulation} The positions of the two nodes are employed to develop a model, $[x_1,y_1, x_2,y_2]$, and the total length of the system is 
\begin{equation}
l = 2l_{1}+l_{2},
\end{equation}
and the uniaxial compression is defined as $\delta l$.
Based on the geometry, the instantaneous length of the left, middle, and right springs are given by,
\begin{equation}
\begin{aligned}
l_L &= \sqrt{w_L^2+y_1^2}, \\
l_R &= \sqrt{w_R^2+y_2^2}, \\
l_C &= \sqrt{w_{C}^2+y_{C}^2},
\end{aligned}
\end{equation}
where
\begin{equation}
\begin{aligned}
w_L &= x_1, \\
w_R &= l - \delta l - x_2, \\
w_{C} &= x_2-x_1, \\
y_{C} &= y_1-y_2. \\
\end{aligned}
\end{equation}
The angle between the two consecutive nodes is formulated as
\begin{equation}
\begin{aligned}
\theta_{L} &= \arctan( \frac{y_1}{w_L}), \\
\theta_{R} &= \arctan(  \frac{y_2}{w_R}), \\
\theta_{LC} &= \theta_L-\Delta\theta, \\
\theta_{RC} &= \theta_R+\Delta\theta,
\end{aligned}
\end{equation}
with
\begin{equation}
\Delta\theta = \arctan( \frac{y_{C}}{w_{C}}).
\end{equation}
Following the previous study \citep{wang2024transient}, we characterize the symmetric and asymmetric responses of the system using two parameters,
\begin{equation}
\begin{aligned}
\theta_S &=\frac{1}{2} (\theta_L+\theta_R), \\
\theta_A & = \frac{1}{2}(\theta_L- \theta_R).
\end{aligned}
\end{equation}
The total energy is formulated as the sum of the stretching and bending energies,
\begin{equation}
\mathcal{U} = \frac{1}{2}k_{s}\Big[(l_L-l_1)^2+(l_R-l_1)^2+(l_C-l_{2})^2\Big] + \frac{1}{2}k_{t}\Big[(\theta_L+\alpha_{L})^2+(\theta_R+\alpha_{R})^2
+\theta_{LC}^2+\theta_{RC}^2\Big],
\end{equation}
{A small imperfection is introduced as a slight difference between the rest angles of the left and right support torsional springs, $\alpha_L = \alpha + \delta \bar{\alpha}/2$ and $\alpha_R = \alpha - \delta \bar{\alpha}/2$, with $\delta \bar{\alpha} = 0.01\,\alpha$.}
The dynamic equilibrium can be  derived  by solving the following equations,
\begin{equation}
\begin{aligned}
m \ddot{x}_{1} + \frac{\partial \mathcal{U}} {\partial{x}_{1}} &= 0, \\
m \ddot{y}_{1} + \frac{\partial \mathcal{U}} {\partial{y}_{1}} &=  0,  \\
m \ddot{x}_{2} + \frac{\partial \mathcal{U}} {\partial{x}_{2}}  &=  0, \\
m \ddot{y}_{2} + \frac{\partial \mathcal{U}} {\partial{y}_{2}} & = 0.
\end{aligned}
\end{equation}
For a prescribed end displacement, the equilibrium configurations are obtained by solving the stationarity of the total elastic energy for the nodal coordinates $\mathbf{q}_{\mathrm{toy}}=[x_{1},y_{1},x_{2},y_{2}]$. Stability is then assessed from the local stiffness matrix, defined as the Hessian of the energy with respect to $\mathbf{q}_{\mathrm{toy}}$: stable equilibria have positive eigenvalues, whereas bifurcation points are identified when the minimum eigenvalue approaches zero. The nodal coordinates are finally transformed into the symmetric and asymmetric coordinates, $\theta_S$ and $\theta_A$, to distinguish symmetry-preserving and symmetry-breaking responses \citep{wang2024transient,giudici2025transient}.

\paragraph{Non-dimensionalization} Next, we rescale all the parameters and make the problem dimensionless.
The normalized physical parameters are used,
\begin{equation}
\mu= \frac {\delta l} { \alpha^2 l},  \; \psi_S= \frac{\theta_S} {\alpha}, \; \psi_A= \frac {\theta_A} {\alpha}, \; \Lambda = \frac{ 3k_{t}}  {\alpha^2 k_{s} l_1^2}, \; L_{2} = \frac {l_{2}} {l_{1}}, \; T = t \sqrt{\frac{k_{s}}{m}}
\end{equation}

\paragraph{Static solution} By using the small-angle approximation, we can get the analytical solution of the equilibrium state \citep{wang2024transient}.
First of all, for the symmetric solution, $\psi_A = 0$, we can solve for $\psi_S$, 
\begin{equation}
\begin{aligned}
    \psi_S^{N} &= -\frac{\sqrt[3]{2} \left(a_3 \mu-a_2\right)}{\sqrt[3]{\sqrt{108 a_4^3
   \left(a_2-a_3 \mu\right){}^3+729 a_1^2 a_4^4}+27 a_1
   a_4^2}}\\
   &\quad-\frac{\sqrt[3]{\sqrt{108 a_4^3 \left(a_2-a_3 \mu\right){}^3+729 a_1^2
   a_4^4}+27 a_1 a_4^2}}{3 \sqrt[3]{2} a_4},\\
   \psi_{S}^I &= - \frac{\left(1+i \sqrt{3}\right) \left(a_2-a_3 \mu\right)}{2^{2/3}
   \sqrt[3]{\sqrt{108 a_4^3 \left(a_2-a_3 \mu\right){}^3+729 a_1^2 a_4^4}+27 a_1
   a_4^2}} \\
   &\quad+\frac{\left(1-i \sqrt{3}\right) \sqrt[3]{\sqrt{108 a_4^3 \left(a_2-a_3
   \mu\right){}^3+729 a_1^2 a_4^4}+27 a_1 a_4^2}}{6 \sqrt[3]{2} a_4},\\
   \psi_{S}^U & = - \frac{\left(1-i \sqrt{3}\right) \left(a_2-a_3 \mu\right)}{2^{2/3}
   \sqrt[3]{\sqrt{108 a_4^3 \left(a_2-a_3 \mu\right){}^3+729 a_1^2 a_4^4}+27 a_1
   a_4^2}} \\
   &\quad+\frac{\left(1+i \sqrt{3}\right) \sqrt[3]{\sqrt{108 a_4^3 \left(a_2-a_3
   \mu\right){}^3+729 a_1^2 a_4^4}+27 a_1 a_4^2}}{6 \sqrt[3]{2} a_4},
\end{aligned}
\end{equation}
where
\begin{equation}
   a_1= \frac{2}{3} \Lambda, \;
   a_2 = \frac{4}{3} \Lambda, \;
   a_3 = \frac{2}{3} (2+L_{2}),\;
   a_4 =\frac{2}{3}.
\end{equation}
Here, $\psi_S^{I}$ and $\psi_S^{N}$ denote the inverted and natural states of the truss, respectively, which are analogous to the plus state $\mathcal{C}^{P}$ and minus state $\mathcal{C}^{M}$ in the folded-ribbon system. The intermediate branch $\psi_S^{U}$ denotes the unstable state connecting these two symmetric configurations.
The critical saddle-node bifurcation point is obtained from the condition $\psi_{S}^I 
 = \psi_{S}^U $,
\begin{equation}
    \psi_S^{\mathrm{SN}}=\left(\frac{\Lambda}{2}\right)^{1/3}.
\end{equation}
Secondly, for the asymmetric solution, $\psi_A \neq 0$, the solution is \citep{wang2024transient}, 
\begin{equation}
\psi_S^{A}= \frac{1}{4}L_{2} \left(L_{2}+2\right),
\end{equation}
and, thus, the critical pitchfork bifurcation occurs at
\begin{equation}
\psi_S^{\mathrm{PF}}= \frac{1}{4}L_{2} \left(L_{2}+2\right).
\end{equation}
On one side, when $\psi_S^{\mathrm{PF}}>\psi_S^{\mathrm{SN}}$, the pitchfork bifurcation is encountered before the saddle-node bifurcation during the release process. The system therefore loses symmetry before snap-through, leading to an asymmetric transition, as shown in Fig.~\ref{fig:toy_model}(b). On the other side, when $\psi_S^{\mathrm{SN}}>\psi_S^{\mathrm{PF}}$, the saddle-node bifurcation is encountered first. In this case, the system loses stability while remaining on the symmetric branch, resulting in a symmetric snap-through transition, as shown in Fig.~\ref{fig:toy_model}(c).

The critical value of $L_2$ separating these two bifurcation-ordering regimes is obtained by solving $\psi_S^{\mathrm{SN}}=\psi_S^{\mathrm{PF}}$, which gives
\begin{equation}
L_{2}^{*} = 2  \sqrt{  \left(\frac{\Lambda}{2} \right)^{1/3} + \frac{1}{4}} - 1.
\end{equation}
This critical value is shown in Fig.~\ref{fig:toy_model}(d). Note that in all plots in Fig.~\ref{fig:toy_model}, we use $\alpha=0.1$ and $\Lambda=0.1$, yielding $L_2^{*}=0.5728$. Thus, varying $L_2$ across this critical value reverses the ordering of the pitchfork and saddle-node bifurcations and switches the predicted snap-through pathway between asymmetric and symmetric modes.

\paragraph{Dynamic validation} 
Finally, we perform dynamic simulations of the reduced-order model to confirm the bifurcation-ordering prediction and to resolve the transient snap-through process. The bifurcation parameter is swept slowly by prescribing a small loading rate $\dot{\mu}$, such that $1-\mu=\dot{\mu}T$, with the previously defined small imperfection to trigger branch selection during symmetry breaking. In the simulations, we use $\dot{\mu}=0.001$, which is sufficiently slow to approximate quasi-static loading while still capturing the dynamic transition between equilibrium states.

\begin{figure}[!h]
    \centering
    \includegraphics[width=0.85\textwidth]{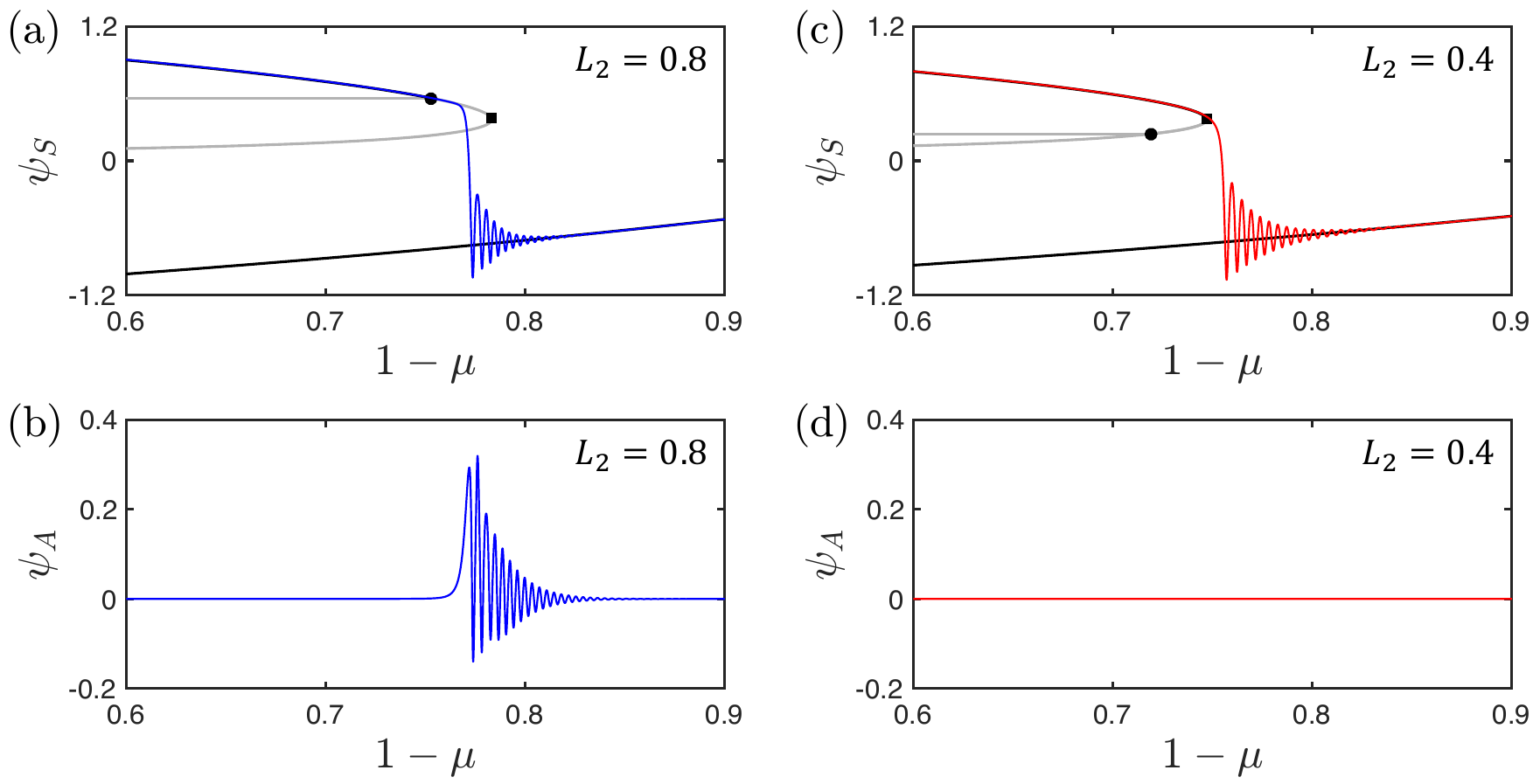}
    \caption{
    Dynamic snap-through responses of the reduced-order double-mass von Mises truss model.
    (a,b) Asymmetric snap-through for $L_2=0.8$.
    (a) Dynamic evolution of the symmetric coordinate $\psi_S$ as a function of the control parameter $1-\mu$, superimposed on the static equilibrium branches. The pitchfork bifurcation is encountered before the saddle-node bifurcation, leading to asymmetric snap-through.
    (b) Corresponding evolution of the asymmetric coordinate $\psi_A$, showing a pronounced transient growth during snap-through.
    (c,d) Symmetric snap-through for $L_2=0.4$.
    (c) Dynamic evolution of $\psi_S$ superimposed on the static equilibrium branches. The saddle-node bifurcation is encountered before the pitchfork bifurcation, triggering symmetric snap-through.
    (d) Corresponding evolution of $\psi_A$, which remains close to zero throughout the transition, confirming the symmetry-preserving response.
    }
    \label{fig:toy_dynamics}
\end{figure}

Figure~\ref{fig:toy_dynamics}(a,b) shows the dynamic response for $L_2=0.8$. In this case, the pitchfork bifurcation is encountered before the saddle-node bifurcation. As $\psi_S$ passes the pitchfork point, the asymmetric coordinate $\psi_A$ grows from zero, indicating that the system loses symmetry before snap-through and transitions through an asymmetric pathway.
{Near the instability, the initial departure from an unstable equilibrium is expected to exhibit locally exponential growth under fixed control parameters, as predicted by linear stability analysis. Under the present slowly varying loading, however, the instantaneous growth rate evolves with the control parameter, and the transient response need not follow a single exponential law.}
Figure~\ref{fig:toy_dynamics}(c,d) shows the corresponding response for $L_2=0.4$. Here, the saddle-node bifurcation is encountered first, and $\psi_A$ remains close to zero throughout the transition. The system therefore snaps while preserving symmetry, consistent with the saddle-node-first ordering predicted by the static analysis.
These dynamic simulations provide time-dependent validation of the static bifurcation analysis, confirming that the relative ordering of the pitchfork and saddle-node bifurcations controls whether snap-through proceeds through an asymmetric or symmetric pathway.

\section{Video}
\label{sec:AppendixVideo}

We provide six videos as supplementary material to illustrate our dynamic results.

\bibliographystyle{elsarticle-harv}
\bibliography{paper}

\end{document}